\documentclass{svjour3}                     
\smartqed  
\usepackage{graphicx}
\usepackage{mathptmx}
\usepackage{tablefootnote}
\usepackage{soul}
\newenvironment{itemlist}{
\begin{list}{}{\setlength{\leftmargin}{10mm}\setlength{\parsep}{0mm}
\setlength{\itemsep}{3mm}} }{\end{list}}
\journalname{Experimental Astronomy}
\begin{document}

\title{The {\it Herschel}-PACS photometer calibration
}
\subtitle{Point-source flux calibration for scan maps}

\titlerunning{PACS Photometer Flux Calibration}        

\author{Zoltan Balog\and
        Thomas M\"uller\and
        Markus Nielbock\and
        Bruno Altieri\and
        Ulrich Klaas\and
        Joris Blommaert\and
        Hendrik Linz\and
        Dieter Lutz\and
        Attila Mo\'or\and
        Nicolas Billot\and
        Marc Sauvage\and
        Koryo Okumura}


\institute{Z. Balog \at
              Max-Planck-Institut f\"ur Astronomie, K\"onigstuhl 17, D-69117 Heidelberg Germany \\
              Tel.: +49 6221 528-354\\
              \email{balog@mpia.de}           
           \and
           T. M\"uller, D. Lutz \at
           Max-Planck-Institut f\"ur Extraterrestrische Physik, Garching, Germany
           \and
           M. Nielbock, U. Klaas, H. Linz \at
           Max-Planck-Institut f\"ur Astronomie, Heidelberg, Germany
           \and
           B. Altieri \at
           ESAC, Madrid, Spain
           \and
           J.A.D.L. Blommaert \at
           Instituut voor Sterrenkunde, K.U.Leuven, Leuven, Belgium
           \and
           A. Mo\'or \at
           Konkoly Observatory, Budapest, Hungary
           \and
           N. Billot \at
           IRAM, Granada, Spain
           \and
           M. Sauvage, K. Okumura \at
           CEA, Saclay, France
}

\date{Received: date / Accepted: date}

\maketitle

\begin{abstract}
This paper provides an overview of the PACS photometer flux calibration concept, in particular for the
principal observation mode, the scan map. The absolute flux calibration is tied to the photospheric models
of five fiducial stellar standards ($\alpha$ Boo, $\alpha$ Cet, $\alpha$ Tau, $\beta$ And, 
$\gamma$ Dra). The data processing steps to arrive at a consistent
and homogeneous calibration are outlined. In the current state the relative photometric accuracy is $\sim$2\% in all bands. Starting from the present calibration status, the characterization and correction for instrumental effects affecting the relative calibration accuracy is described and an outlook for the final achievable calibration numbers is given. After including all the correction for the instrumental effects, the relative photometric calibration
accuracy (repeatability) will be as good as 0.5\% in the blue and green band and 2\% in the red band.
This excellent calibration starts to reveal possible inconsistencies between the models of the K-type  and the M-type stellar calibrators.
The absolute calibration accuracy is therefore mainly limited by the 5\% uncertainty of the celestial
standard models in all three bands. The PACS bolometer response was extremely stable over the entire Herschel mission and a single, time-independent response calibration file is sufficient for the processing and calibration of the science observations. The dedicated measurements of the internal calibration sources were needed only to characterize secondary effects. No aging effects of the bolometer or the filters have been found. Also, we found no signs of filter leaks. The PACS photometric system is very well characterized with a constant energy spectrum $\nu F_{\nu} = \lambda F_{\lambda} = const$ as a reference. Colour corrections for a wide range of sources SEDs are determined and tabulated.

\keywords{	Herschel Space Observatory \and
	PACS \and
	Far-infrared \and 
	Instrumentation \and 
	Calibration \and
	Scan-map}
\end{abstract}

\section{Introduction}
\label{intro}
The Photodetector Array Camera and Spectrograph (PACS)~\cite{Refpoglitsch10}
on board the Herschel Space Observatory~\cite{Refpilbratt10} provides
photometric imaging and integral field spectroscopy capabilities for the
far-infrared wavelength regime. The PACS photometer imaging unit is a dual
band imaging camera. It allows the simultaneous observation in two bands, 
either in the combination 70\,$\mu$m$(blue)+$160\,$\mu$m(red) or 
100\,$\mu$m$(green)+$160\,$\mu$m(red), where 70 or 100\,$\mu$m bands are selected by a 
filter wheel. The camera contains two bolometer detector arrays with 
64$\times$32 pixels (blue array) and 32$\times$16 pixels (red array), 
respectively, providing an instantaneous field-of-view of 
3.5$\times$1.75\,arcmin$^2$. The detector arrays are made of 8 and 2 filled 
bolometer matrices with 16$\times$16 pixels each, respectively.
The arrays operate at $\approx$285\,mK provided by a $^3$He sorption cooler
with a hold time of 57.8\,h, if the bolometers are biased all the time after
the cooler recycling.  

The principal science observation mode with the PACS photometer is the scan
map with the telescope scanning along parallel legs covering the map area\footnote{
(for further reference see  {\tt http://herschel.esac.esa.int/Docs/PACS/html/pacs\_om.html})}.
The nominal scan map speed is 20$^{\prime\prime}$/s. In most cases both an in-scan and a 
cross-scan map were taken in order to reduce the striping in the combined maps and for getting a better handle on extended structures.
Two flavors of scan maps were used:
\begin{itemize}
\item[1)] Large scan maps with scan angle orientations of usually 45 and
          135$^o$. The scan leg separation is freely selectable ranging
          from relatively wide separations ensuring homogeneous coverage
          of large fields to narrow separations for high redundancy\footnote{{\tt http://herschel.esac.esa.int/twiki/pub/Public/PacsAotReleaseNotes//PACS\_ScanMap\_ReleaseNote\_23Feb2010.pdf}}.
\item[2)] Mini scan maps for dedicated point/compact source photometry
          have only extensions in the order of about 3\,arcmin, scan angle
          orientations of 70 and 110$^o$ and narrow scan leg separation
          of a few arcsec. This mode is also used for the flux calibration\footnote{\tt http://herschel.esac.esa.int/twiki/pub/Public/PacsCalibrationWeb/PhotMiniScan\_ReleaseNote\_20101112.pdf}.
\end{itemize}  

An alternative observation mode, the chop/nod point source photometry,
has been maintained over the full mission for flux calibration consistency
checking and in support of observatory pointing calibration. Its flux
calibration is described in this issue~\cite{Refnielbock13}. 

\section{Basic Photometric Calibration Strategy}
\label{sec:2}
The physical quantity which determines the bolometer's absolute photometric 
calibration is its responsivity R. R, in units [V/W], is the ratio of the measured output signal U$_{\rm sig}$ produced by the
  infalling far-infrared radiation power F$_{\rm photband}$, hence
\begin{equation}
U_{\rm sig} = R \times F_{\rm photband} 
\end{equation} 

With a source SED, conventionally expressed as f$_{\nu,s}(\lambda)$ [Jy] in
the FIR, and the relation \begin{math} 
f_{\lambda} = \frac{c}{\lambda^2}~f_{\nu} \end{math}, the flux in a PACS 
photometer band can be expressed as \\
\begin{equation}
F_{\rm photband} [W] = T~A~ 
                       \int_{\lambda_{1}}^{\lambda_{2}} \frac{c}{\lambda^2}
                       ~f_{\nu,s}(\lambda)~S(\lambda)~d\lambda
\end{equation} \\
with T being the product of reflection losses by the optical mirrors, A being
the effective telescope area and S($\lambda$) being the relative photometer 
system response of the PACS photometer band.

For the PACS photometer, the convention is, that the flux density at the
reference wavelength f$_{\nu,s}(\lambda_{0})$ is determined for the reference
SED f$_{\nu,1} = \nu^{-1}$, i.e. \begin{math} f_{\nu,1}(\lambda) = 
\frac{\nu_{0}}{\nu}~f_{\nu,1}(\lambda_{0}) = 
\frac{\lambda}{\lambda_{0}}~f_{\nu,1}(\lambda_{0})\end{math}, hence

\begin{equation}
F_{\rm photband} [W] = T~A~f_{\nu,1}(\lambda_{0})~\frac{c}{\lambda_{0}}
                       \int_{\lambda_{1}}^{\lambda_{2}} \frac{1}{\lambda}
                       ~S(\lambda)~d\lambda
\end{equation} \\
\begin{equation}
f_{\nu,1}(\lambda_{0}) = \frac{F_{\rm photband}}{T~A~\frac{c}{\lambda_{0}} 
                         \int_{\lambda_{1}}^{\lambda_{2}} \frac{1}{\lambda}
                         ~S(\lambda)~d\lambda} = F_{\rm photband}~C_{\rm conv}
\end{equation} \\
with \\
\begin{equation}
C_{\rm conv} = \frac{1}{T~A~\frac{c}{\lambda_{0}} 
               \int_{\lambda_{1}}^{\lambda_{2}} \frac{1}{\lambda}
                     ~S(\lambda)~d\lambda} 
             = \frac{1}{T~A~\Delta\nu_{0}}
\end{equation} \\
and effective bandwidth \\
\begin{equation}
\Delta\nu_{0} = \frac{c}{\lambda_{0}} \int_{\lambda_{1}}^{\lambda_{2}} 
                \frac{1}{\lambda}~S(\lambda)~d\lambda
\end{equation}

If the source SED f$_{\nu,s}(\lambda)$ is known, the true photometer source
flux is determined as \\
\begin{equation}
f_{\nu,s}(\lambda_{0}) = \frac{f_{\nu,1}(\lambda_{0})}{K_{cc}}
\end{equation} \\
with K$_{cc}$ being the appropriate color correction factor as described in \cite{Refpoglitsch10} and in the
document ``PACS Photometer Passbands and Colour Correction Factors for various 
source SEDs'', PICC-ME-TN-038\footnote{{\tt http://herschel.esac.esa.int/twiki/pub/Public/PacsCalibrationWeb/cc\_report\_v1.pdf}} 

This means that \\
\begin{equation}
~\\
R \propto \frac{1}{F_{\rm photband}} \propto \frac{C_{\rm
    conv}}{f_{\nu,1}(\lambda_{0})} \propto \frac{1}{T \times A \times
  \Delta\nu_{0} \times f_{\nu,s}(\lambda_{0})
  \times K_{cc}}
~\\
\end{equation} \\
hence the responsivity R can be calibrated, too, by flux density measurements
of celestial standards.

The responsivity of the bolometer arrays depends on their operating temperature, voltage bias, and optical loading (see Sec. 6). R was thoroughly characterized in the lab (cf.~\cite{Refbillot06}), and the ground-based calibration was used to estimate the initial in-flight responsivity at the beginning of the mission when the telecope foreground emission had stabilized.
For in-flight calibration
only a relative update in the form of \\
\begin{equation}
\frac{R_{new}}{R_{old}} = \frac{f_{\nu,s}(\lambda_{0})}{f_{\nu,standard model}(\lambda_{0})}
\end{equation} \\
was necessary, by using the ratio of the measured flux
to the celestial standard star model flux, when calibrating 
$f_{\nu,s}(\lambda_{0})$ with R$_{old}$.

In practice, the update of R did not rely on one single
flux measurement alone, but on a set of measurements on
several standards to minimize systematic effects \\
\begin{equation}
\frac{R_{new}}{R_{old}} = \frac{1}{n} \sum_{i=1}^{n} \frac{f_{\nu,s,i}(\lambda_{0})}{f_{\nu,standard model,i}(\lambda_{0})}
\end{equation} \\
so that \begin{math} \frac{1}{n} \sum_{i=1}^{n}
\frac{f_{\nu,s,i}(\lambda_{0})}{f_{\nu,standard model,i}(\lambda_{0})} = 1 \end{math}
by applying R$_{new}$. 

This scheme applies, when more standard star measurements
become available, when additional systematic instrument
effects altering the signal level are corrected for, when changes
in the encircled energy fraction alter the resulting flux
(see below) or when new standard star models become available.
 
Note, that R is a global value for the whole detector array.
Individual pixel-to-pixels variations are accounted for
by the flat-field ff$_{pix}$.

R is not a constant. It depends on temperature and the total flux load,
hence R = R(T, B$_{total flux}$), as will be shown later.

The standard evaluation procedure for flux calibration
measurements is aperture photometery. Hence, not the
the total source flux is measured, but only the fraction
inside the measurement aperture. For the derivation of the
total measured source flux a correction for the encircled 
energy fraction 
(EEF\footnote{\tt http://herschel.esac.esa.int/twiki/pub/Public/PacsCalibrationWeb/bolopsf\_20.pdf}, see Table \ref{tbl:EEF}) 
inside the aperture must be performed: \\
\begin{equation}
f_{\nu,1}(\lambda_{0}) = \frac{f_{\nu,1}^{aperture}(\lambda_{0})}{f_{EEF}}
\end{equation} \\

\section{Photometric Calibration Standards}
\label{sec3}
As prime flux reference, models of the photospheric emission
of late type giants are used~\cite{Refdehaes11}. This type
of stars was already used as absolute calibrators for earlier 
IR space missions (e.g.\ IRAS:~\cite{Refrieke85}, ISO:~\cite{Refcohen01}, 
Spitzer:~\cite{Refgordon07}). The flux regime they cover
is inside the linear flux behaviour of the PACS bolometers.
Ideally, stars of different stellar types would be used to prevent systematic uncertainties from the modeling. Unfortunately only the late type giants are bright enough at far infrared wavelengths to be observed at high signal to noise. One star, Sirius ($\alpha$ CMa), was the only A type exception, but unfortunately had to be dropped from our set of primary flux calibrators as will be discussed in \ref{sec3.2}
Models of the atmospheres of the giant planets Uranus and 
Neptune ~\cite{Refmoreno12} are equally accurate, however 
these sources are already in the non-linear flux regime
of PACS. Asteroids are about to be established as independent
prime FIR flux calibrators (~\cite{Refmueller13}, this volume).
The five fiducial stellar standards $\alpha$Boo, $\alpha$Cet,
$\alpha$Tau, $\beta$And and $\gamma$Dra were observed repeatedly
during the mission for absolute calibration and to monitor
the system stability (for this $\gamma$Dra, which was visible
during the whole mission, was observed on a monthly basis). 
We also monitored the photometric system using the Calibration Blocks (CalBlocks). They track the response several times a day throughout the mission (at least one Calblock per OBSID) and show the high stability of the photometric system with high accuracy (see \cite{Moor13}). 
 
\subsection{Summary of mini scan-map observations}
\label{sec:3.1}
All measurements were taken as part of one of the following calibration
programmes: ``RPPhotFlux\_321B", ``RPPhotFlux\_324B", ``RPPhotFlux\_321D",
``PVPhotAOTVal\_514P". All observations were taken in medium scan
speed with 20$^{\prime \prime}$/s, in ``high gain" and only one single
repetition of each scan-map. All measurements were preceded by a 30-s calibration measurement
 on the PACS internal calibration sources (see \cite{Moor13})The detailed information on the observation of the individual fiducial stars can be found in Tables \ref{tbl:aBoo}, \ref{tbl:aCet}, \ref{tbl:aTau}, \ref{tbl:bAnd}, and \ref{tbl:gDra}. 
 All observations taken with the PACS Photometer from Operational Day (OD)\#1375 (16 February 2013) onwards are affected by a serious anomaly detected in one of the two red channel subarrays (matrix 9). Data taken in ODs\#1375 and \#1376 from this subarray were found to be uniformly saturated. Recovery activities taking place in subsequent days were unsuccessful. Further observations taken in OD\#1380 confirmed the problem. It was also confirmed that the other red channel subarray (matrix 10), was working correctly. This results essentially in a degradation of a factor 1.4 in sensitivity and reduced spatial coverage of all the images taken in the 160 micron band, ranging from some reduction of the map area, to a loss of up to one third of the map area in the most extreme cases. Altogether six calibration observations (taken after OD\#1376) were affected by this anomaly. Fortunately the layout of our observations and the flux level of the sources ensured that the loss in sensitivity did not influence the photometry.

\subsection{Model fluxes}
\label{sec3.2}
The theoretical spectra of candidate stellar calibrators in the far infrared were generated using the MARCS stellar atmosphere code (\cite{Refgustafsson75}, \cite{Refgustafsson03}) and the TURBOSPECTRUM synthetic spectrum code (\cite{Refplez92}) and are presented in \cite{Refdehaes11}. The stellar parameters and their uncertainties were derived by \cite{Refdecin03}. The line lists used in the spectrum calculations and the model uncertainties are discussed in \cite{Refdecin07}, where it is estimated that the uncertainty of the models in the PACS wavelength range is $\approx 5$\%. The absolute flux calibration is based on Selby's (\cite{Refselby88}) K-band photometry, the zero-point is determined on the basis of an ideal 'Vega', i.e. the K-band photometry of Vega is corrected for a flux excess of 1.29\% (cf. \cite{Refabsil06}). The determined Selby K-band zeropoint is $4.0517\times10^{-10}$ W/m$^2$/$\mu$m. This and more detailed information on the individual stars can be found in the headers of the fits files containing the theoretical spectra used for the PACS flux calibration. The files can be found at ftp.ster.kuleuven.be/dist/pacs/calsources/ or ftp.sciops.esa.int and will be provided in the Herschel archive. 

As the stars can show excess flux in the far infrared wavelengths, due to either debris disks for early type dwarfs or in the case of late type giants, by a chromosphere or ionised wind, it is important to investigate whether we can rule out the existence of such excesses for our candidate calibrators. 
\cite{Refdehaes11} present observations in the sub-mm up to cm wavelength range for nine late type giant stars, of which seven were in our original list of candidate calibrators. These observations were made in preparation of the Herschel mission and performed on several ground based telescopes. Also Sirius, a A1V star, was studied, but not included in the \cite{Refdehaes11} paper as this only considered K and M giants. Although in seven out of nine studied giants an excess was detected, it was found that for the eight stars, selected as prime flux calibrators for the PACS photometer, the excess only started beyond the PACS wavelength range.  

Based on early analysis of Herschel observations of potential calibration stars
 we decided to use only the following 5 fiducial stars for the final analysis: $\beta$~And, $\alpha$~Cet, $\alpha$~Tau, $\alpha$~Boo, $\gamma$~Dra. The three other potential calibrators were discarded ($\alpha$~CMa, $\beta$~Peg, $\beta$~UMi). In the case of these three stars the fluxes deviated from the models by more than 10\% in at least one PACS band. For $\beta$~UMi no Selby K-band photometry was available for the absolute calibration and only a less accurate ($\sim 10\%$) Johnson K band from \cite{Refducati02} could be used. \cite{Refprice04} showed that $\beta$ Peg is variable in the mid-IR bands by about 10\%.  $\alpha$ CMa shows about 20\% excess at 160 $\mu$m. The underlying cause of the discrepancy is still unclear and is under investigation.
 
The monochromatic flux densities based on at 70.0, 100.0 and 160.0\,$\mu$m for the five finally selected stars are
given in Table~\ref{tbl:tbl2}. 

\begin{table}
\centering
\caption{Information on the selected fiducial stars. Monochromatic flux densities at
70.0, 100.0 and 160.0\,$\mu$m are given. The stellar  temperatures are taken from \cite{Refdecin03} and the fluxes are based on models published in \cite{Refdehaes11}.}
  \label{tbl:tbl2}
{\footnotesize
\begin{tabular}{rrrlllllrrrr}
\noalign{\smallskip}
\hline
\noalign{\smallskip}
     &        &       &              &             &               &          & Temp    & \multicolumn{3}{c}{Model flux [mJy]} \\
 HR  &  HD    & HIP   & ID            & RA (J2000)    & Dec (J2000)   & SpType  & [K]       & 70\,$\mu$m  & 100\,$\mu$m & 160\,$\mu$m \\
\noalign{\smallskip}
\hline
\noalign{\smallskip}
 337 &   6860 &  5447 &  $\beta$ And & 01:09:43.9236 & +35:37:14.008 & M0III    &  3880     &  5594	& 2737     &  1062 \\
 911 &  18884 & 14135 & $\alpha$ Cet       & 03:02:16.8    & +04:05:24.0   & M1.5IIIa &  3740     &  4889	& 2393     &   928 \\
1457 &  29139 & 21421 & $\alpha$ Tau   & 04:35:55.2387 & +16:30:33.485 & K5III    &  3850     & 14131	& 6909     &  2677 \\
5340 & 124897 & 69673 & $\alpha$ Boo   & 14:15:39.6720 & +19:10:56.677 & K1.5III  &  4320     & 15434	& 7509     &  2891 \\
6705 & 164058 & 87833 & $\gamma$ Dra      & 17:56:36.3699 & +51:29:20.022 & K5III    &  3960     &  3283	& 1604     &   621 \\
\noalign{\smallskip}
\hline
\noalign{\smallskip}
\end{tabular}}
\footnotetext{Footnote}
\end{table}

\section{Determination of flux densities}
\label{sec4}
\subsection{Data reduction and calibration}
\label{sec:4.1}

\subsubsection{Pre-processing}
\label{sec4.1.1}
The data reduction was done with software version "hcss.dp.pacs-11.0.2600", calibration tree No. 58 and the following steps, parameters and calibration file versions:
\begin{itemlist}
\item[$\bullet$] Flagging of bad pixels (badPixelMask: FM, 6)
\item[$\bullet$] Flagging of saturated pixels (clSaturationLimits: FM, 1; satLimits: FM, 2)
\item[$\bullet$] Conversion digital units to Volts
\item[$\bullet$] Adding of pointing and time information 
                (we used the updated pointing product for observations taken between OD 320 and OD 761)
\item[$\bullet$] Response calibration (responsivity: FM, 7)
\item[$\bullet$] Flat fielding (flatField: FM, 4)
\item[$\bullet$] Frame selection based on scan speed ($\pm$10\% of the nominal speed) \\
({\tt frames = filterOnScanSpeed(frames, limit=10.0, copy=None)})
\item[$\bullet$] Second level deglitching 
\item[$\bullet$] Save the observation context into a fits-file (including still all frames)
\end{itemlist}

\subsubsection{Post-processing}
\label{sec4.1.2}
\begin{figure}
\includegraphics[width=0.49\textwidth, angle=90]{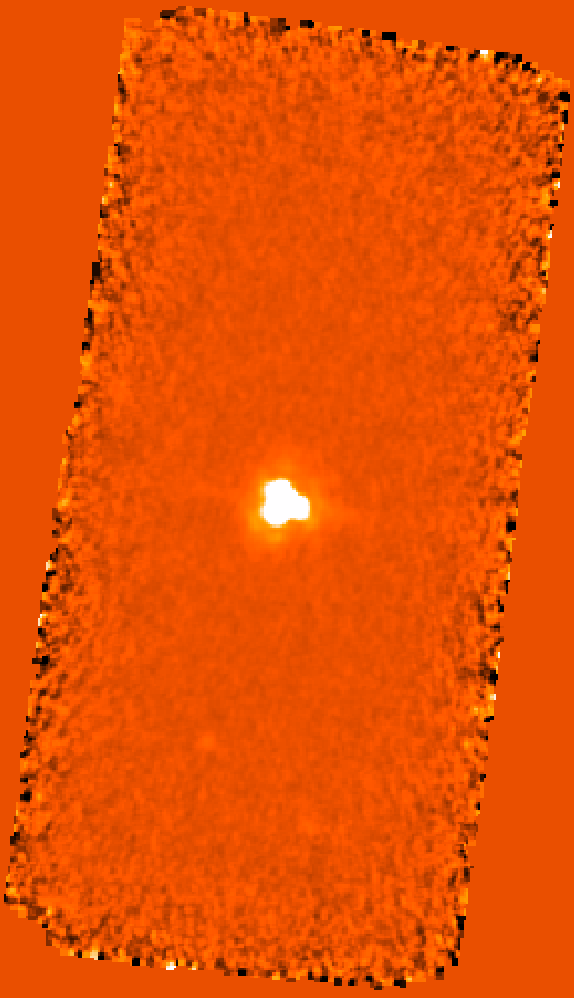}
\caption{Mini scanmap of $\alpha$ Tau  obtained at 70 $\mu$m as a typical example for a flux calibration measurement.}
\label{fig:alphaTau}    
\end{figure}

The purpose of post processing is to eliminate the 1/f noise of the detector using high-pass filtering of the data. The process includes two steps.
The purpose of the first step is to locate and mask the source using the level2 product and a circular mask with a size of 25$^{\prime\prime}$, while the second steps 
includes the high-pass filtering of the data.\\
\begin{itemlist}
\item[$\bullet$] Set on-target flag for all frames to 'true'
\item[$\bullet$] Set a circular mask around the position of the source (r=25$^{\prime\prime}$)
\item[$\bullet$] High-pass filtering with a hp-filter width of 15, 20, 35 in blue, green
                 and red band respectively (adjusted to the FWHM of the corresponding PSFs, see Section 4.1.3 for details)
\item[$\bullet$] Merging frames ({\tt join}) of both scan directions (usually 70$^{\circ}$
                 and 110$^{\circ}$ scan angles in instrument frame: approx.\ along the diagonals of the bolometer array)
\item[$\bullet$] Frame selection based on scan speed ($\pm$10\% of the nominal speed)
\item[$\bullet$] Final projection of all data with {\tt photProject()}, using the default pixel fraction
                 (pixfrac = 1.0) and reduced pixel sizes of 1.1$^{\prime \prime}$, 1.4$^{\prime \prime}$, 2.1$^{\prime \prime}$
\item[$\bullet$] Save final map as fits-file (see Fig.~\ref{fig:alphaTau})

\end{itemlist}

\subsubsection{Determination of the high-pass filter width}
\label{sec4.1.3}
We determined the high-pass filter width in each band to account for the difference in their FWHM of the PSF and to make sure that the mask is large enough so that the high-pass filtering does not remove flux from the source. All the data used in the calibration was taken at 20$^{\prime \prime}$/s.  The bolometer data are taken with 40\,Hz with an onboard averaging of 4 frames in both channels in the PACS prime mode leading to a data rate of 10\,Hz in the downlink. The FWHM of a point-source is about 5.6$^{\prime \prime}$ in the blue band, 6.8$^{\prime \prime}$ in the green band
                 and 11.3$^{\prime \prime}$ (average values for 20$^{\prime \prime}$/s scan speed), so in a signal time-line for a given pixel a central hit of the source has therefore the
                 following width: 5.6$^{\prime \prime}$ / 20$^{\prime \prime}$/s = 0.28\,s or 2.8 frames in blue,
		 6.8$^{\prime \prime}$ / 20$^{\prime \prime}$/s = 0.34\,s or 3.4 frames in green,
		 11.3$^{\prime \prime}$ / 20$^{\prime \prime}$/s = 0.565\,s or 5.65 frames in red. The values 15, 20, and 35 for the high-pass-filter parameter correspond to 31, 41, and 71 frames per filter width (2 $\times$ hpf-parameter + 1). Considering the scanning speed of 20$^{\prime\prime}$/s and a 10-Hz sampling, these frame numbers translate
to 62$^{\prime\prime}$, 82$^{\prime\prime}$, and 142$^{\prime\prime}$ filter width at 70, 100, and 160 $\mu$m,
respectively. The ratios between FWHM and high-pass filter width are 2.8/31=0.09 in blue, 3.4/41=0.08 in green and 5.65/71=0.08 in red. These ratios are very similar in the three bands and at a very conservative level so that the high-pass filtering is not affecting the source flux with the appropriate masking.

\clearpage
\subsection{Photometry}
\label{sec4.2}
\subsubsection{Source flux determination}
\label{sec4.2.1}
We determined the flux using aperture photometry with aperture sizes of 12$^{\prime \prime}$, 12$^{\prime \prime}$, 22$^{\prime \prime}$ in blue, green, and red, respectively (centering the aperture on the flux peak). These values are based on an analysis of the influence of the high-pass filter width on aperture photometry \cite{Pope13}. It shows that a small aperture is better, as long as the aperture correction is applied. The larger the aperture, the more the photometry will depend on the filter width. These apertures are chosen such that the uncertainty because of the high-pass filter width is less than 1\%. This approach has the advantage that the results can be relatively independent of high-pass filter width. Note, that at lower flux levels one should select significantly smaller aperture sizes for better S/N.

\begin{figure}
\includegraphics[width=0.49\textwidth]{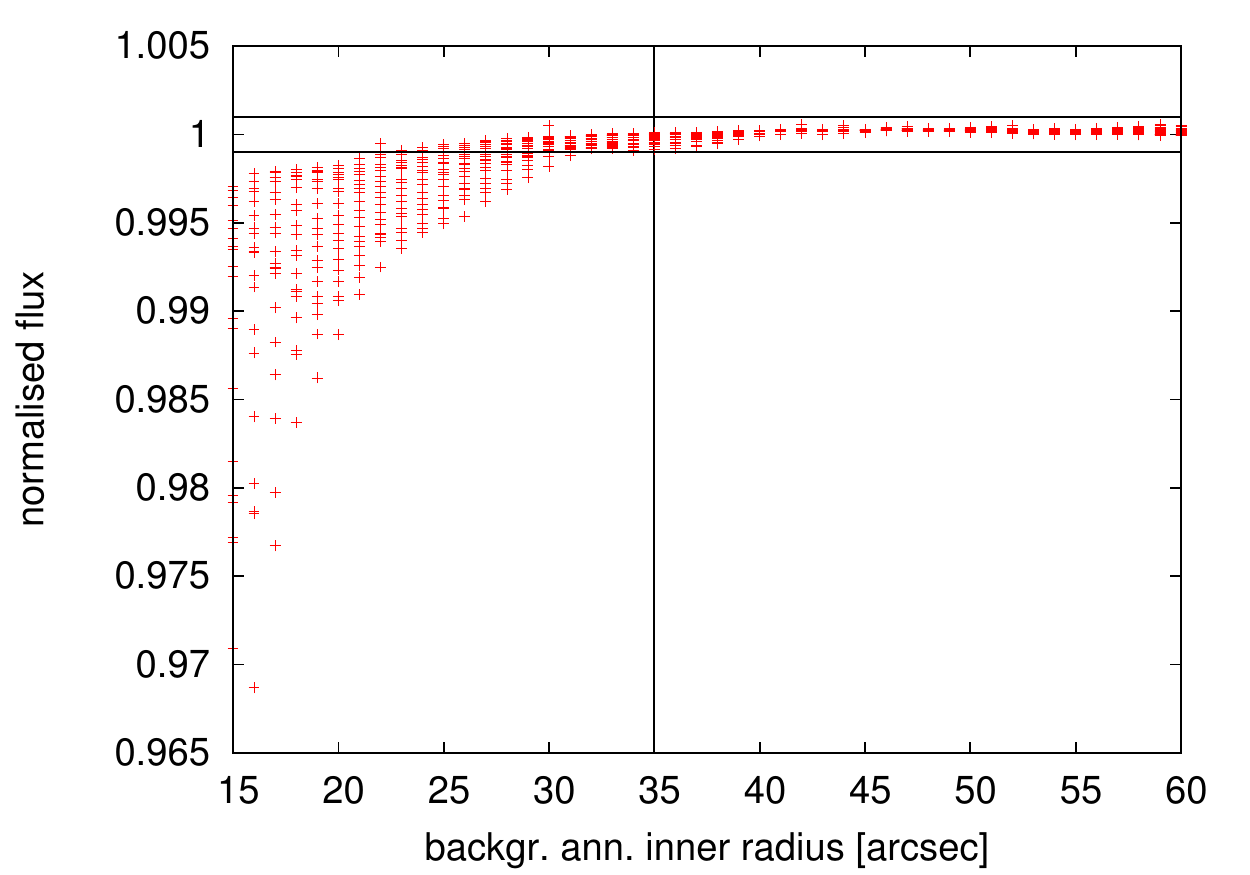}
\caption{Photometry of $\alpha$ Boo using different background annuli. The fluxes are normalized to the value measured using the annulus between 60$^{\prime \prime}$ and 80$^{\prime \prime}$. The vertical line shows the inner radius of the background annulus used for our photometry, while the horizontal line designate the $\pm$0.1\% limit.}
\label{fig:background}  
\end{figure}

For background subtraction, we selected an annulus between 35$^{\prime\prime}$ and 45$^{\prime\prime}$. We applied aperture corrections of 0.802, 0.776, and 0.817 for blue, green and red filters respectively (see Table \ref{tbl:EEF}\footnote{see {\tt http://herschel.esac.esa.int/twiki/pub/Public/PacsCalibrationWeb/bolopsf\_20.pdf} for an in depth analysis of the PACS PSF and the calculation of the encircled energy fractions} ), to account for the flux outside the aperture. These values are based on very large maps (extending to 15$^{\prime}$ from a bright source) indicating that about 10\% more flux is in the outermost wings of the PSFs beyond 60$^{\prime \prime}$. Although the wings of the PSF extending far outside the outer limit of our background annulus there is no leftover flux from the PSF affecting our photometry because the high-pass filtering effectively removes the source flux from the image outside the boundary of our mask (r=25$^{\prime\prime}$) while it leaves the flux unaffected inside. Detailed investigation of the effect of the background annulus on the final photometry of the standard stars show that our measurements are affected by less than 0.1\% by the leftover flux of the PSF wings within the background annulus. As an illustration, Fig. \ref{fig:background} shows the photometry of $\alpha$ Boo (the brightest of our five fiducial stars; see Table \ref{tbl:tbl2}) at 70 $\mu$m using different background annuli. The horizontal axis corresponds to the inner edge while the different points at the same inner radius value refer to different outer radii. To guide the reader we show the inner radius of the annulus used for the photometry of the standards (vertical line) and the $\pm$0.1\% limits (horizontal lines) with respect to the photometry using the outermost annulus (between 60$^{\prime \prime}$ and 80$^{\prime \prime}$). It is clear from the figure that the placement and the size of the background annulus does not affect the photometry by more than 0.1\% outside of the 30$^{\prime \prime}$ inner radius. At smaller inner radii however the effect is noticeable and needs to be corrected. The PACS ICC is currently working on the new calibration file that corrects this effect in HIPE.

Note, that the previous values, connected to earlier versions of PSFs (and assuming that there is no flux beyond 60$^{\prime \prime}$) and to the same aperture sizes, were: 0.886 (blue band, 12$^{\prime \prime}$ aperture radius), 0.866 (green band, 12$^{\prime \prime}$ aperture radius), and 0.916 (red band, 22$^{\prime \prime}$ aperture radius). These values were correct for responsivity calibration file (FM,5), but are not correct for later versions\footnote{see {\tt http://herschel.esac.esa.int/twiki/pub/Public/PacsCalibrationWeb/PhotMiniScan\_ReleaseNote\_20101112.pdf}}. 
                      
\subsubsection{Source flux uncertainty determination}
\label{sec4.2.2}
We determined the flux uncertainty of the calibration source measurements in two different ways.
\begin{itemlist}
\item[1] By performing aperture photometry on the background annulus. We used the same aperture sizes as for the source photometry. We distributed six apertures evenly along the background annulus, measured the flux without background subtraction and determined the standard deviation of the measured values after aperture correction. 
  
\item[2] By estimating the noise inside the annulus used for background subtraction in all three bands (signal r.m.s.\ inside the selected sky annulus). Note that the noise per pixel in the final maps depends on the selected pixel size, the
high-pass filter width and also the ``drop size". We only used one set of
fixed values for the high-pass filtering (15 in blue, 20 in green and 35
in red band) and the default drop size (pixfrac = 1.0), the so called correlated
noise in the final maps is therefore a consequence of the
regridding the data from the pixel values of the detector array into maps
projected on the sky. This modifies the measured SNR
(signal-to-noise ratio). In particular, if small map pixels are selected,
the measured SNR of the integrated flux can appear to be much higher than the true value. The 
correction for the correlated noise (rms-noise values have to be divided by these factors) amounts to:
\begin{enumerate}
                 \item[-] blue band:  $1.00 \times (pixsize/3.2)^{1.78}$; here: $1.00 \times (1.1/3.2)^{1.78} = 0.1495$ \\
                 ($\rightarrow$ noise increases by a factor of 6.7)
                 \item[-] green band: $1.01 \times (pixsize/3.2)^{1.70}$; here: $1.01 \times (1.4/3.2)^{1.70} = 0.2477$ \\
                 ($\rightarrow$ noise increases by a factor of 4.0)
                 \item[-] red band:   $1.02 \times (pixsize/6.4)^{1.51}$; here: $1.02 \times (2.1/6.4)^{1.51} = 0.1896$ \\
                 ($\rightarrow$ noise increases by a factor of 5.3)
            \end{enumerate}
The correction applied to the photometric uncertainty was established in
the following way. Using a template observation, the rms of the
pixel-to-pixel flux variations was calculated for each fully calibrated
Level 1 detector frame. The ratio between the theoretical final
photometric uncertainty by assuming a Gaussian error propagation of the
Level 1 frames and the value measured in the final Level 2 was obtained
for varying map pixel sizes and default drizzling drop sizes. This set
of ratios is used for correcting the photometric uncertainties for the
unaccounted correlated noise.
Finally we determined the error of the source flux using the following formula: 
$$corrected,\,\, r.m.s.\,\,values\,\, \times {\sqrt{(number\,\, of\,\,map\,\,pixel\,\,inside\,\,the\,\,specified\,\,aperture)} \over {aperture\,\,correction}}$$
\end{itemlist}

\begin{table}[h!tb]
\begin{center}
\caption{Encircled Energy Fraction (EEF) as a function of circular aperture radius for
the three PACS filter bands. Valid for responsivity calibration file version FM, 7. See {\tt http://herschel.esac.esa.int/twiki/pub/Public/PacsCalibrationWeb/bolopsf\_20.pdf} for an in depth analysis of the PACS PSF and the calculation of the encircled energy fractions}
\label{tbl:EEF}
\begin{tabular}{|cccc|cccc|}
\noalign{\smallskip}
\hline
radius & \multicolumn{3}{c|}{Encircled Energy Fraction} &radius & \multicolumn{3}{c|}{Encircled Energy Fraction} \\ 
$[ \,\,\prime\prime\,\, ]$ & blue & green  & red&$[ \,\,\prime\prime\,\, ]$ & blue & green  & red \\
\hline
2 &  0.192 &  0.141 &  0.060 &  32 &  0.896 &  0.891 &  0.861\\
3 &  0.353 &  0.278 &  0.127 &  33 &  0.898 &  0.893 &  0.864\\
4 &  0.487 &  0.413 &  0.209 &  34 &  0.900 &  0.895 &  0.867\\
5 &  0.577 &  0.521 &  0.298 &  35 &  0.902 &  0.896 &  0.870\\
6 &  0.637 &  0.595 &  0.384 &  36 &  0.904 &  0.898 &  0.873\\
7 &  0.681 &  0.641 &  0.461 &  37 &  0.905 &  0.900 &  0.876\\
8 &  0.719 &  0.673 &  0.527 &  38 &  0.907 &  0.902 &  0.879\\
9 &  0.751 &  0.700 &  0.579 &  39 &  0.908 &  0.903 &  0.882\\
10 &  0.774 &  0.727 &  0.619 &  40 &  0.910 &  0.905 &  0.884\\
11 &  0.791 &  0.753 &  0.649 &  41 &  0.911 &  0.906 &  0.886\\
12 &  0.802 &  0.776 &  0.673 &  42 &  0.913 &  0.908 &  0.888\\
13 &  0.812 &  0.795 &  0.694 &  43 &  0.914 &  0.909 &  0.890\\
14 &  0.820 &  0.808 &  0.712 &  44 &  0.916 &  0.910 &  0.892\\
15 &  0.829 &  0.818 &  0.729 &  45 &  0.917 &  0.912 &  0.894\\
16 &  0.837 &  0.826 &  0.746 &  46 &  0.919 &  0.913 &  0.896\\
17 &  0.845 &  0.832 &  0.761 &  47 &  0.920 &  0.914 &  0.897\\
18 &  0.852 &  0.837 &  0.776 &  48 &  0.921 &  0.915 &  0.899\\
19 &  0.858 &  0.842 &  0.789 &  49 &  0.922 &  0.917 &  0.901\\
20 &  0.863 &  0.847 &  0.800 &  50 &  0.924 &  0.918 &  0.902\\
21 &  0.867 &  0.852 &  0.809 &  51 &  0.925 &  0.919 &  0.904\\
22 &  0.870 &  0.857 &  0.817 &  52 &  0.926 &  0.920 &  0.905\\
23 &  0.874 &  0.863 &  0.824 &  53 &  0.927 &  0.921 &  0.906\\
24 &  0.877 &  0.867 &  0.830 &  54 &  0.929 &  0.922 &  0.908\\
25 &  0.880 &  0.872 &  0.835 &  55 &  0.930 &  0.923 &  0.909\\
26 &  0.883 &  0.876 &  0.839 &  56 &  0.931 &  0.924 &  0.910\\
27 &  0.885 &  0.879 &  0.843 &  57 &  0.932 &  0.925 &  0.911\\
28 &  0.888 &  0.882 &  0.847 &  58 &  0.933 &  0.926 &  0.912\\
29 &  0.890 &  0.885 &  0.850 &  59 &  0.935 &  0.928 &  0.913\\
30 &  0.892 &  0.887 &  0.854 &  60 &  0.936 &  0.929 &  0.914\\
31 &  0.894 &  0.889 &  0.857 &  61 &  0.937 &  0.929 &  0.915\\
\hline
\noalign{\smallskip}
\end{tabular}
\end{center}
\end{table}

Almost always the first method resulted in larger values so we accepted the values from this method as the final error estimate. Note, that the source flux uncertainties in case of the fiducial stars are very small. SNR values are well above 100. The error bars throughout this report are therefore in most cases smaller than the symbol sizes.

\subsubsection{Colour correction}
\label{sec4.2.3}
The colour correction values for the fiducial stars (1.016, 1.033,
1.074) were derived from a 4000 K black body, very close
to the effective temperature of the stars (see Table.~\ref{tbl:tbl2}),
ranging between 3740\,K and 4320\,K. We repeated now the
calculation using the official model template files (see section \ref{sec3.2}) and could confirm the tabulated values
in the blue and red band. In the green band we found a difference
of 0.1\% (1.034) when using the full stellar templates instead of
a 4000 K black body. No difference between the K- and M-giants
are seen on the per mille level. In the final error budget this
small deviation can be neglected and the colour correction is
not contributing to the systematic differences between K- and
M-giants seen in the calibrated flux densities of the 5 fiducial
stars.

The following colour correction factors have been used to obtain
monochromatic flux densities and uncertainties at 70.0, 100.0, 160.0\,$\mu$m:\\

\begin{center}
\begin{tabular}{lll}
\noalign{\smallskip}
\hline
\noalign{\smallskip}
70.0\,$\mu$m & 100.0\,$\mu$m & 160.0\,$\mu$m \\
\noalign{\smallskip}
\hline
\noalign{\smallskip}
1.016 & 1.033 & 1.074 \\
\noalign{\smallskip}
\hline
\noalign{\smallskip}
\end{tabular}
\end{center}

\section{End of mission calibration status}
\label{sec5}
The results for the fiducial stars we obtained based on the responsivity calibration file ``FM, 7" are listed in Table \ref{tbl:oldstd}.
\begin{table}[h!tb]
\begin{center}
\caption{Observed and calibrated (``FM, 7") monochromatic flux densities
at 70, 100, 160\,$\mu$m divided by the corresponding current model predictions
for all 5 fiducial stars.}
  \label{tbl:oldstd}
\begin{tabular}{lrrrrrrrrr}
\noalign{\smallskip}
\hline
\noalign{\smallskip}
Target  & \multicolumn{3}{c}{blue obs/model} & \multicolumn{3}{c}{green obs/model} & \multicolumn{3}{c}{red obs/model} \\
name    & no.\ & mean & stdev & no.\ & mean & stdev & no.\ & mean & stdev  \\
\noalign{\smallskip}
\hline
\noalign{\smallskip}
  $\beta$ And &  6 & {\bf 1.017} & 0.007 &  6 & {\bf 1.017} & 0.007 &  6 & {\bf 0.991} & 0.010 \\
 $\alpha$ Cet &  7 & {\bf 1.013} & 0.006 &  7 & {\bf 1.013} & 0.005 &  7 & {\bf 1.020} & 0.015 \\
 $\alpha$ Tau &  7 & {\bf 0.968} & 0.013 &  7 & {\bf 0.969} & 0.015 &  8 & {\bf 0.967} & 0.015 \\
 $\alpha$ Boo &  7 & {\bf 0.986} & 0.015 &  7 & {\bf 0.993} & 0.014 &  7 & {\bf 1.001} & 0.021 \\
 $\gamma$ Dra & 54 & {\bf 0.984} & 0.009 &  11 & {\bf 0.990} & 0.010 & 60 & {\bf 1.010} & 0.029 \\
\noalign{\smallskip}
\hline
\noalign{\smallskip}
\multicolumn{2}{r}{\bf mean/stdev} & \multicolumn{2}{c}{\bf 0.994$\pm$0.021} & & \multicolumn{2}{r}{\bf 0.997$\pm$0.019} && \multicolumn{2}{r}{\bf 0.998$\pm$0.020} \\
\noalign{\smallskip}
\hline
\noalign{\smallskip}
\end{tabular}
\end{center}
\end{table}
The derived mean values show that the currently available calibration files with the aperture correction provide a very accurate point source flux calibration. The intrinsic accuracy of the photometry is around 2\% in all three filter bands. Considering the 5\% uncertainty of the models of the fiducial stars, the final photometric calibration uncertainty is less than 7\%.

\begin{figure}
\includegraphics[width=0.49\textwidth]{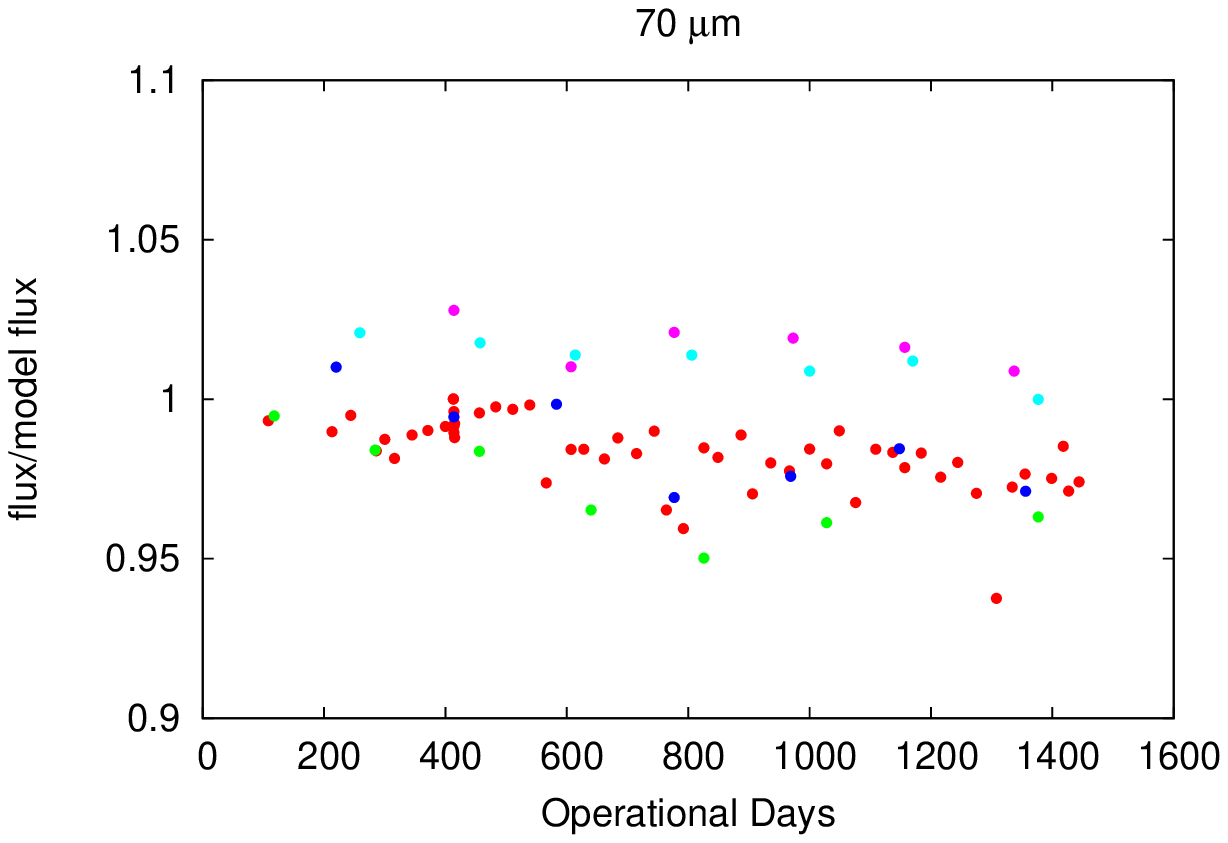}
\includegraphics[width=0.49\textwidth]{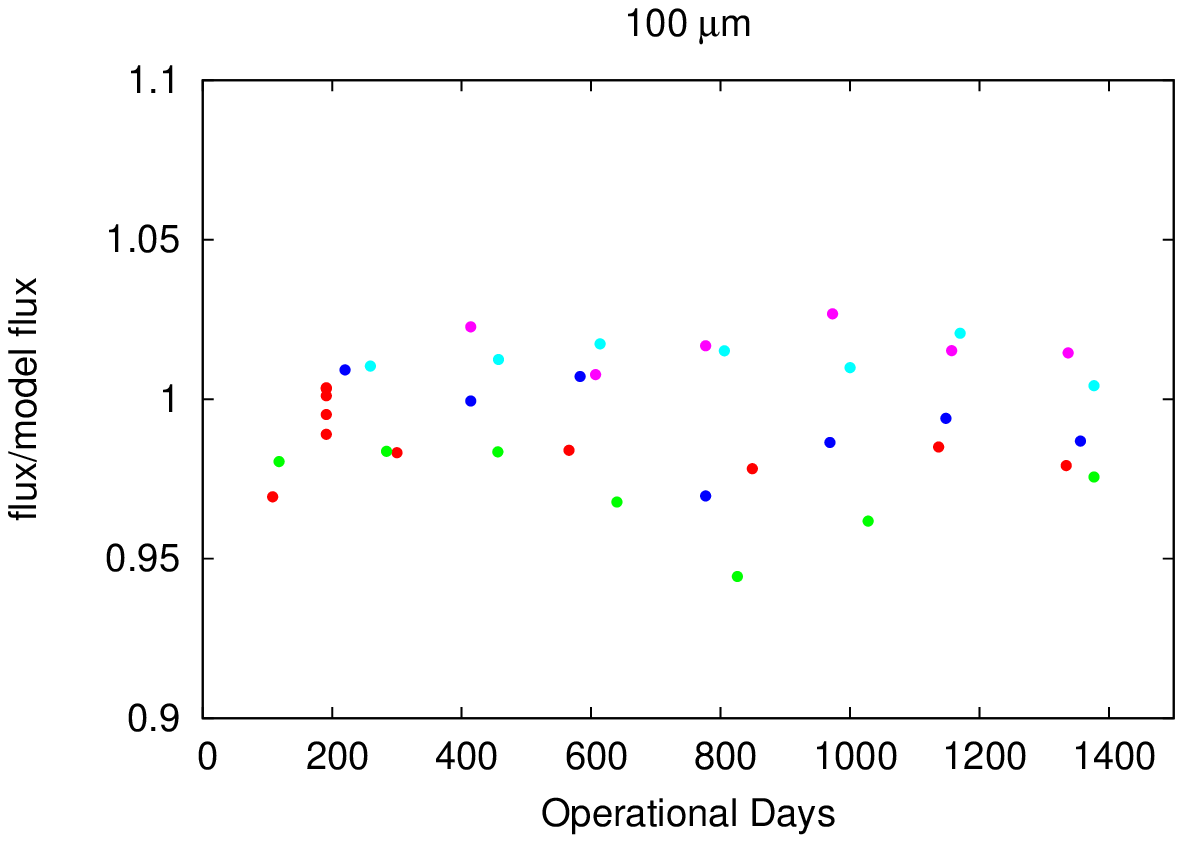}
\includegraphics[width=0.49\textwidth]{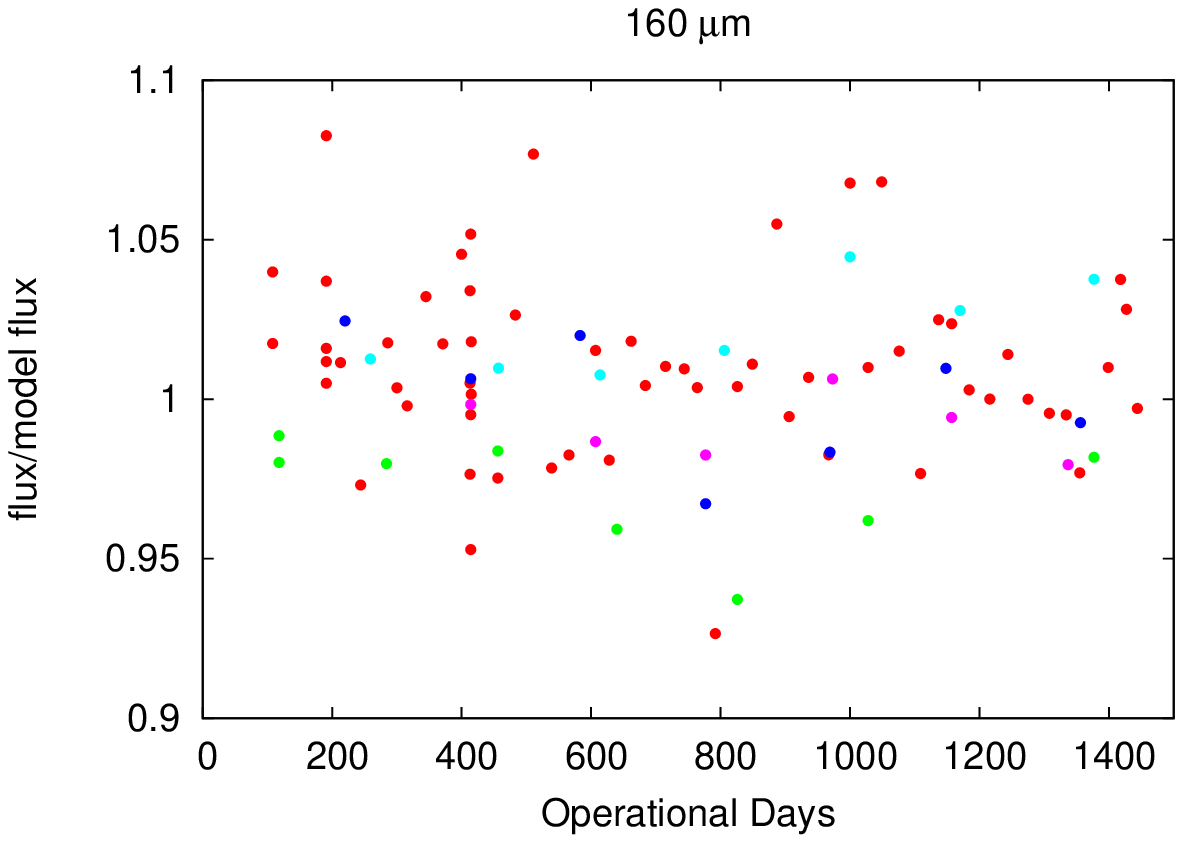}
\caption{Measured over model flux ratios of the five fiducial standards at 70 $\mu$m (upper left), 100 $\mu$m (upper right) and 160 $\mu$m (lower left). The different colours designate different objects. Red: $\gamma$ Dra; green: $\alpha$ Tau; blue: $\alpha$ Boo; magenta: $\beta$ And; cyan: $\alpha$ Cet. We follow this convention throughout the paper.}
\label{fig:oldstd}  
\end{figure}

Figure \ref {fig:oldstd} shows the measured over model flux ratios for all individual measurements at 70 $\mu$m (upper left), 100 $\mu$m (upper right) and 160 $\mu$m (lower left).   There is one measurement of $\gamma$ Dra at 70 $\mu$m that deviates from the locus of the other measurements on the same object. The measurement occurred on OD 1308 when a large pointing offset was introduced by not updating the S/C velocity vector by MOC during the DTCP, so as a precaution we leave this point out of the analysis. 

It is interesting to note that while the individual scatter per source is clearly the largest in the red band the standard deviation of the averages are rather similar. Also in the blue and green bands the fiducial star measurement are systematically different (e.g. the flux ratio of the 2 M-type giants ($\beta$ And and $\alpha$ Cet) are larger by about 3-4\% than the flux ratio of the three K-type giants). This shows that uncertainty of the calibration is dominated by the discrepancy of the model fluxes. 

The numbers in Table \ref{tbl:oldstd} show that the calibrations are equally good at all wavelengths. So the relative photometric accuracy between the different PACS bands is better than 1\%.

\begin{figure}
\includegraphics[width=0.75\textwidth]{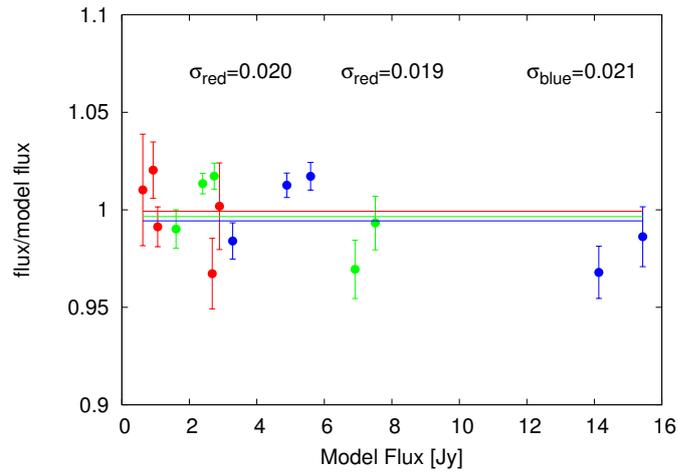}
\caption{Final result of the calibration. Red dots: 160 $\mu$m; green dots: 100 $\mu$m; blue dots: 70 $\mu$m. The red, green and blue lines represent the average values at  160 $\mu$m, 100 $\mu$m, and 70 $\mu$m respectively. The error bars represent the standard deviation of the individual set of measurements (each filter/object combination)}
\label{fig:finalres} 
\end{figure}

Figure \ref{fig:finalres} shows the results summarised  in Table \ref{tbl:oldstd}.

\section{Characterization of instrumental effects affecting the relative calibration accuracy}
\label{sec6}
A closer look at Figure \ref{fig:oldstd} reveals obvious trends in all three bands. There is a long time trend showing that the flux ratios (thus theß measured fluxes) are getting smaller as time progresses. Also there is a variation on shorter timescale.  The intrinsic accuracy of the photometry and the repeatability of the measurements can be increased to large extent by taking into account two issues that are mainly responsible for these variations and affect the flux measured on the array.

\subsection{$^3$He cooler evaporator temperature}
\label{sec6.1}
\begin{figure}
\includegraphics[width=0.49\textwidth]{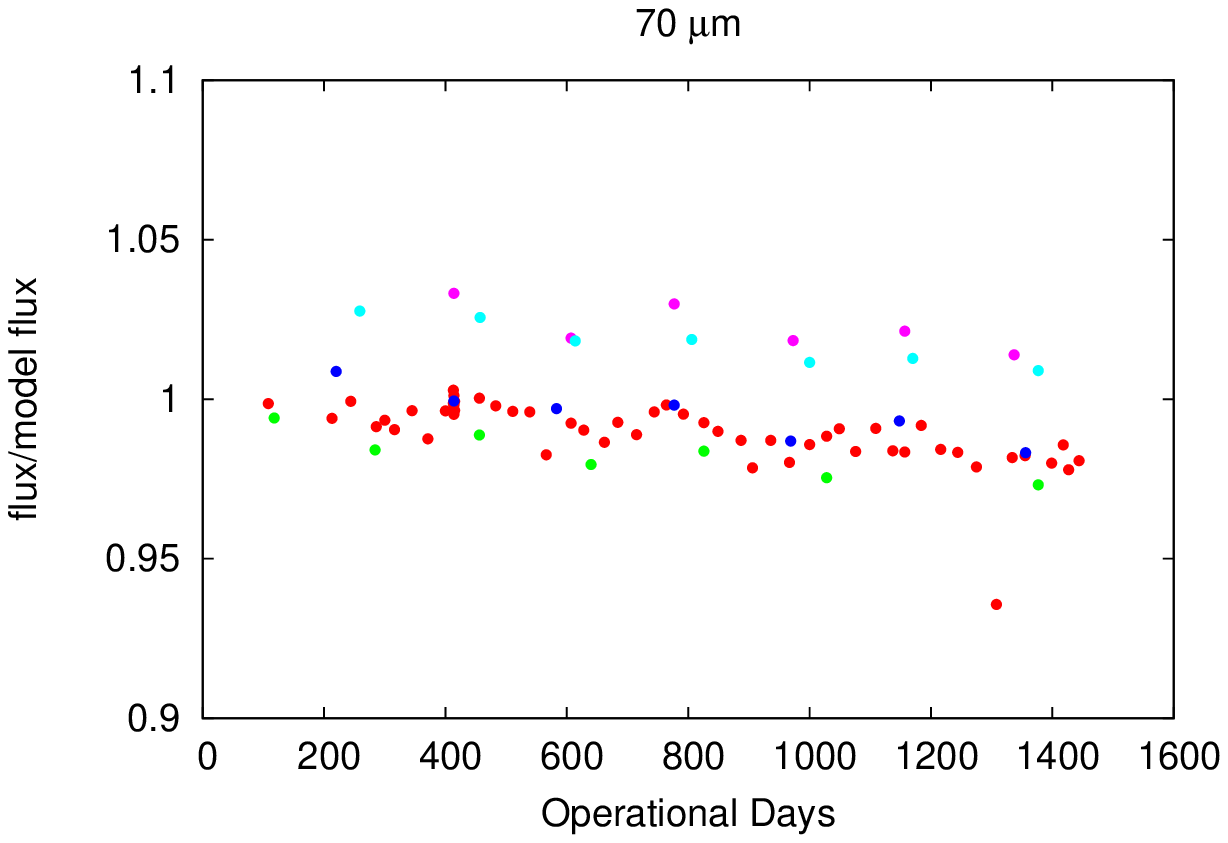}
\includegraphics[width=0.49\textwidth]{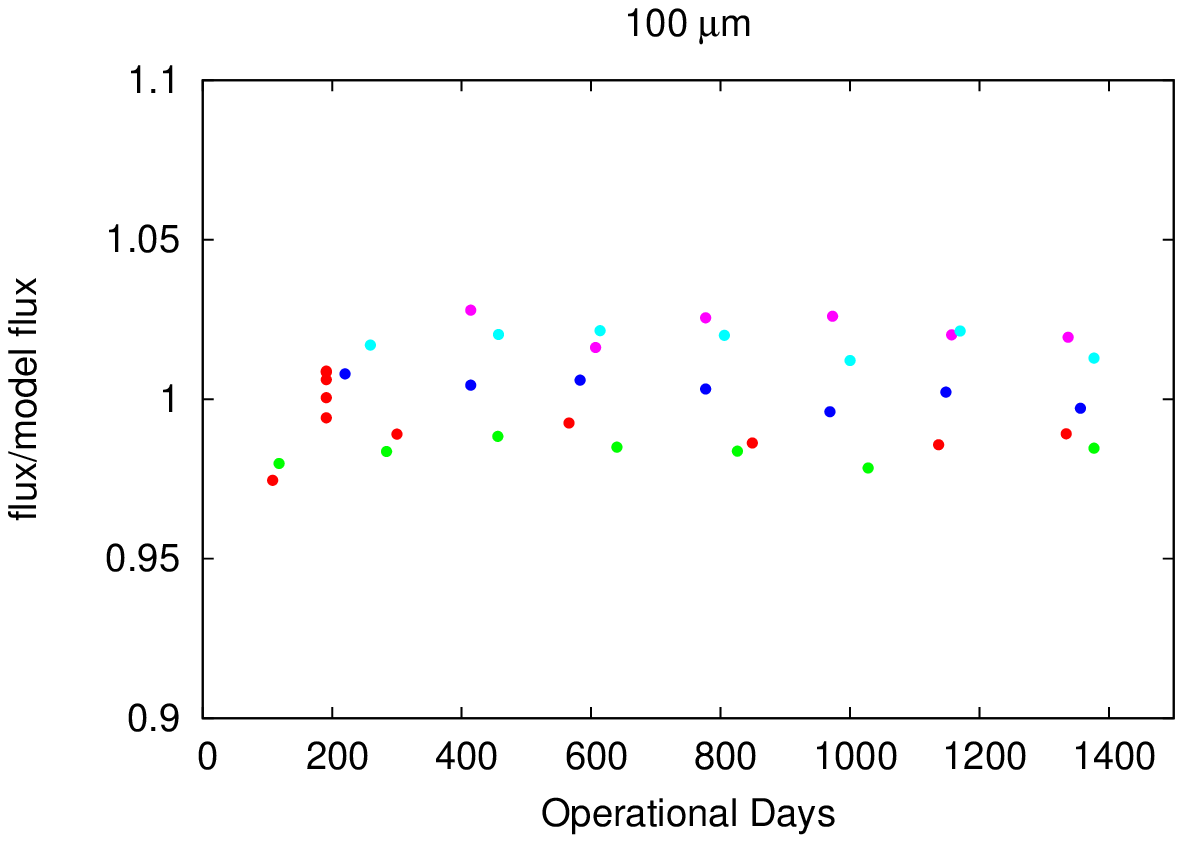}
\includegraphics[width=0.49\textwidth]{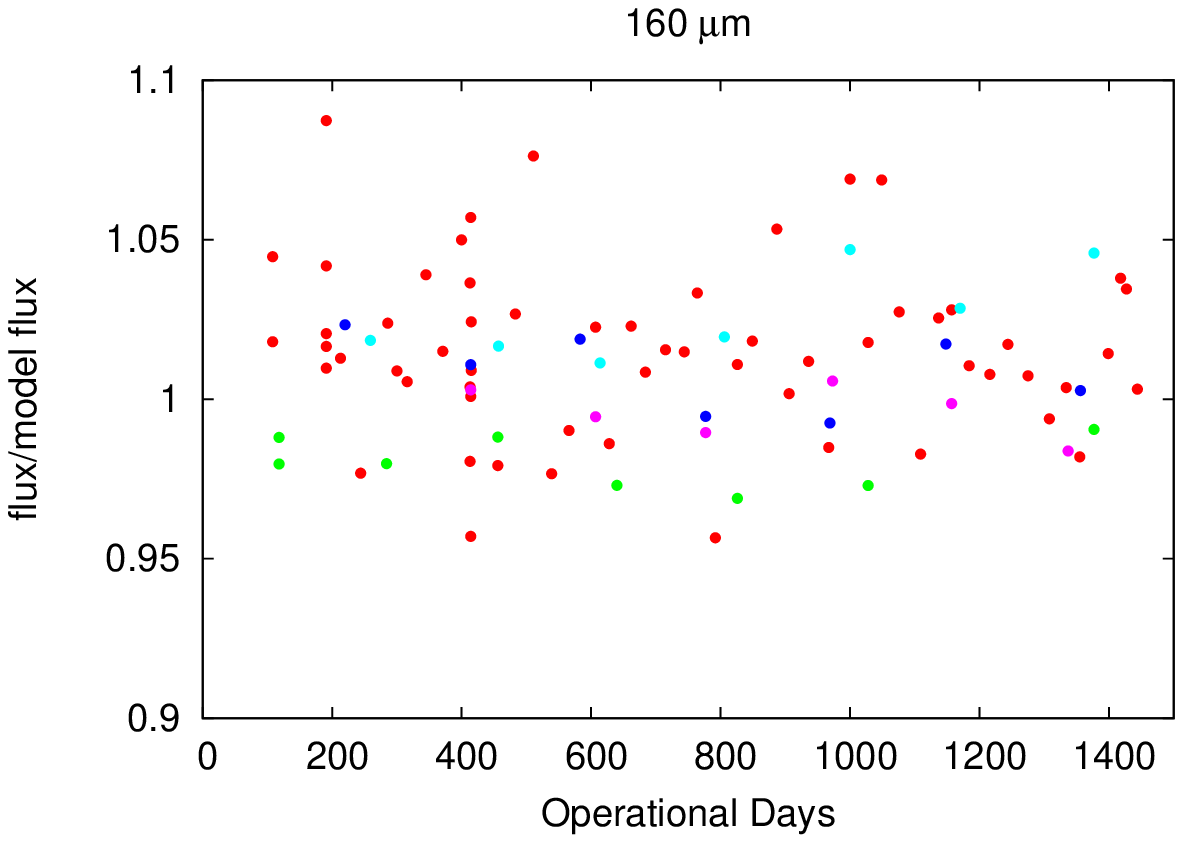}
\caption{Measured over model flux ratios of the five fiducial standards at 70 $\mu$m (upper left), 100 $\mu$m (upper right) and 160 $\mu$m (lower left) after applying the correction for the evaporator temperature.}
\label{fig:oldstd_EVC}
\end{figure}

\cite{Moor13} showed a good correlation between the temperature of the evaporator and the differential calibration block signal (see Fig. 5. of \cite{Moor13}). Based on this relation a pixel-based correction has been established \cite{Moor13}. Applying this correction to the data after calibrating the responsivity, results in more accurate flux measurements. This correction is typically below 1\%. Only 1.2\% of all observations need larger correction. See \cite{Moor13} for details.  The task that performs this correction will be available in HIPE 12.0. 
Figure \ref{fig:oldstd_EVC} shows the same as Figure \ref {fig:oldstd} after applying the correction for the evaporator temperature. The scatter of the individual objects has been greatly reduced that leads to the clear recognition of the second effect: the change of the bolometer response with the telescope flux. 

\subsection{Variation of telescope background}
\label{sec6.2}
\begin{figure}
\includegraphics[width=0.49\textwidth]{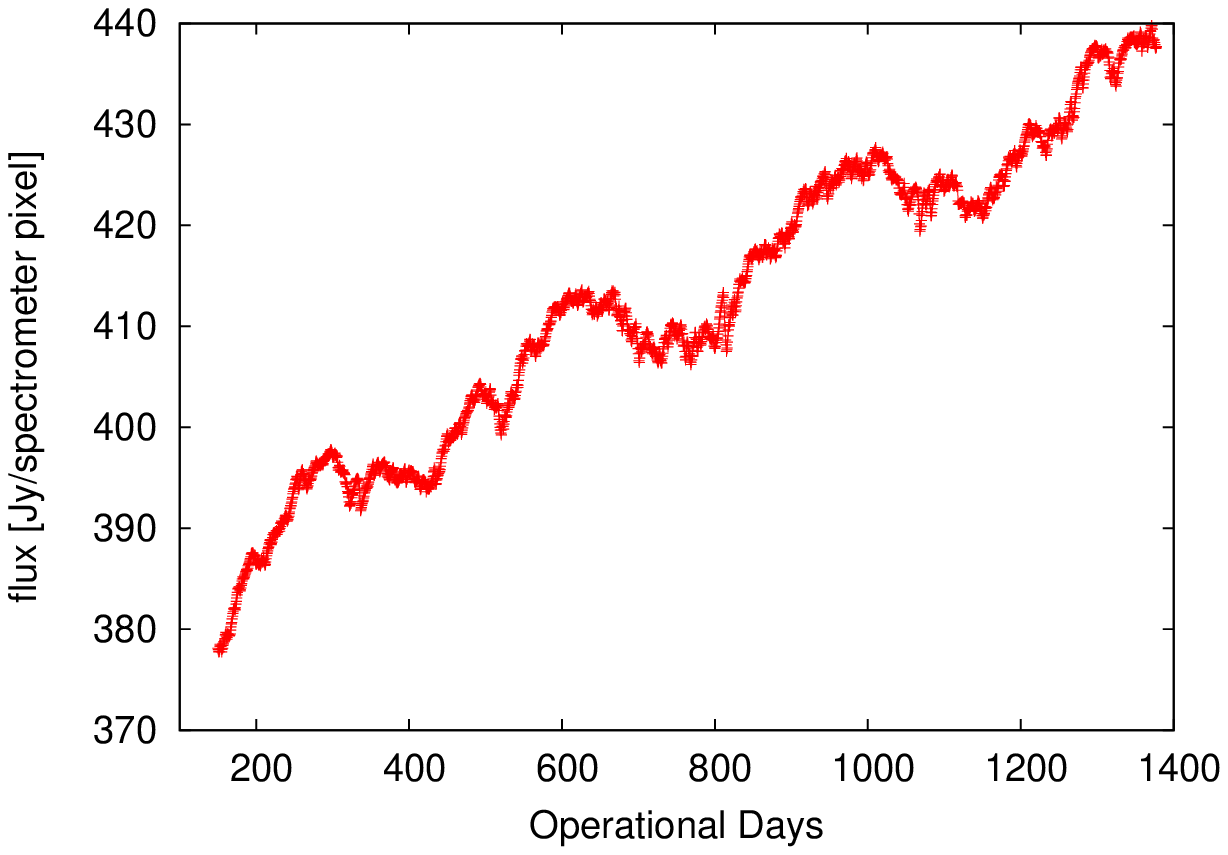}
\includegraphics[width=0.49\textwidth]{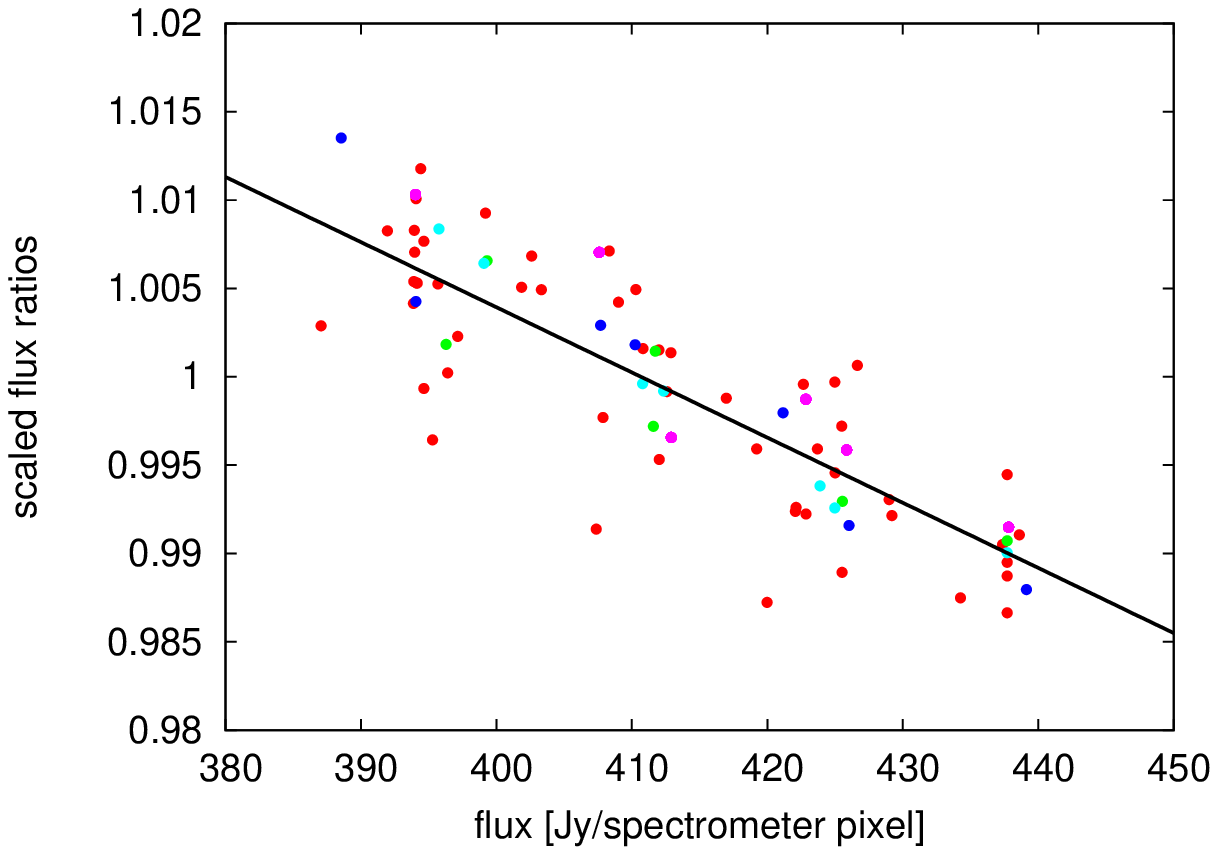}
\caption{The 70 $\mu$m background flux of the main mirror integrated over the beam of a PACS spectrometer pixel (9.4$^{\prime\prime} \times$~9.4$^{\prime\prime}$)as a function of the Operational Days (left panel) and the correlation between the flux ratios of the fiducial stars and the main mirror flux at 70 $\mu$m (right panel)}
\label{fig:telflux} 
\end{figure}

It is a general feature of the bolometers that their responsivity decreases with increasing optical load. Examining the photometric variation of $\gamma$ Dra that was monitored regularly during the mission lead to the realization that the flux showed seasonal changes that correlated well with the calculated telescope flux based on the temperature of the main mirror. It is known that the emission of the main mirror is the largest contributor of the flux impinging on the detector (Figure \ref{fig:telflux} left panel as an example of the 70 $\mu$m flux (Albrecht Poglitsch private communication)) . This calculation was originally provided for the purpose of Telescope Background Normalization of the spectrometer, therefore the unit is {\it Jy/(spectrometer pixel)}).  The curve on Figure \ref{fig:telflux} (left panel) can  be interpreted as annual temperature variations of the primary, where the bumps occur at the orbit perigee, on top of a slowly increasing emissivity of the primary mirror due to e.g. dust deposition on the reflecting surface, surface degradation by cosmic rays. We can derive a linear correlation between the measured flux ratios and the telescope flux (Figure \ref{fig:telflux} right panel.)

The derived functions are the following:
\begin{eqnarray}
f_{blue}(x)=-0.000369*x+1.151418\\
f_{green}(x)=-0.000884*x+1.267293\\
f_{red}(x)=-0.002811*x+1.561422
\end{eqnarray}
 where {\it x} is the telescope flux in the appropriate band. Note that to derive these correlations we need to scale the flux ratios of the fiducial stars to bring them on the same level. The scaling would not influence our results because it does not change the slope of the curves. Since we are interested only in the relative change, we divide the measured flux by $f(x)/f(c)$ where {\it c} is the flux value where the conversion results in no change ($f(x)/f(c) = 1$). The values of {\it c} are 410.65, 302.24 and 199.75 Jy/(spectrometer pixel) in the blue, green and red band respectively. The maximum effect of this correction is around 1.5\%.  
 
\section{Finally achievable calibration accuracies}
\label{sec7}
The result of the calibrations after applying  both the evaporator temperature correction and the telescope flux correction are summarised in Table \ref{tbl:oldstd_EVC_tflux}. 

\begin{table}[h!tb]
\begin{center}
\caption{Observed and calibrated (``FM, 7") monochromatic flux densities
at 70, 100, 160\,$\mu$m divided by the corresponding current model predictions
for all 5 fiducial stars after the evaporator temperature and the telescope flux corrections.}
  \label{tbl:oldstd_EVC_tflux}
\begin{tabular}{lrrrrrrrrr}
\noalign{\smallskip}
\hline
\noalign{\smallskip}
Target  & \multicolumn{3}{c}{blue obs/model} & \multicolumn{3}{c}{green obs/model} & \multicolumn{3}{c}{red obs/model} \\
name    & no.\ & mean & stdev & no.\ & mean & stdev & no.\ & mean & stdev  \\
\noalign{\smallskip}
\hline
\noalign{\smallskip}
  $\beta$ And &  6 & {\bf 1.025} & 0.003 &  6 & {\bf 1.024} & 0.004 &  6 & {\bf 0.997} & 0.008 \\
 $\alpha$ Cet &  6 & {\bf 1.013} & 0.006 &  6 & {\bf 1.019} & 0.003 &  6 & {\bf 1.028} & 0.016 \\
 $\alpha$ Tau &  5 & {\bf 0.981} & 0.002 &  5 & {\bf 0.985} & 0.003 &  5 & {\bf 0.980} & 0.009 \\
 $\alpha$ Boo &  7 & {\bf 0.996} & 0.003 &  7 & {\bf 1.003} & 0.002 &  7 & {\bf 1.009} & 0.011 \\
 $\gamma$ Dra & 49 & {\bf 0.991} & 0.004 &  10 & {\bf 0.994} & 0.006 & 54 & {\bf 1.015} & 0.027 \\
\noalign{\smallskip}
\hline
\noalign{\smallskip}
\multicolumn{2}{r}{\bf mean/stdev} & \multicolumn{2}{c}{\bf 1.003$\pm$0.019} & & \multicolumn{2}{r}{\bf 1.005$\pm$0.017} && \multicolumn{2}{r}{\bf 1.006$\pm$0.018} \\
\noalign{\smallskip}
\hline
\noalign{\smallskip}
\end{tabular}
\end{center}
\end{table}

Although the overall accuracy of the calibration did not improve significantly, the uncertainties of the individual sources were reduced at least by a factor of two in the blue band and 1.5 in the green band to bring the intrinsic accuracy (repeatability) of the individual calibration sources to around 0.5\%. In the red band the corrections do not improve the scatter of the individual sources significantly. Here the conservative estimate of the repeatability remained around 2\%.

\begin{figure}
\includegraphics[width=0.49\textwidth]{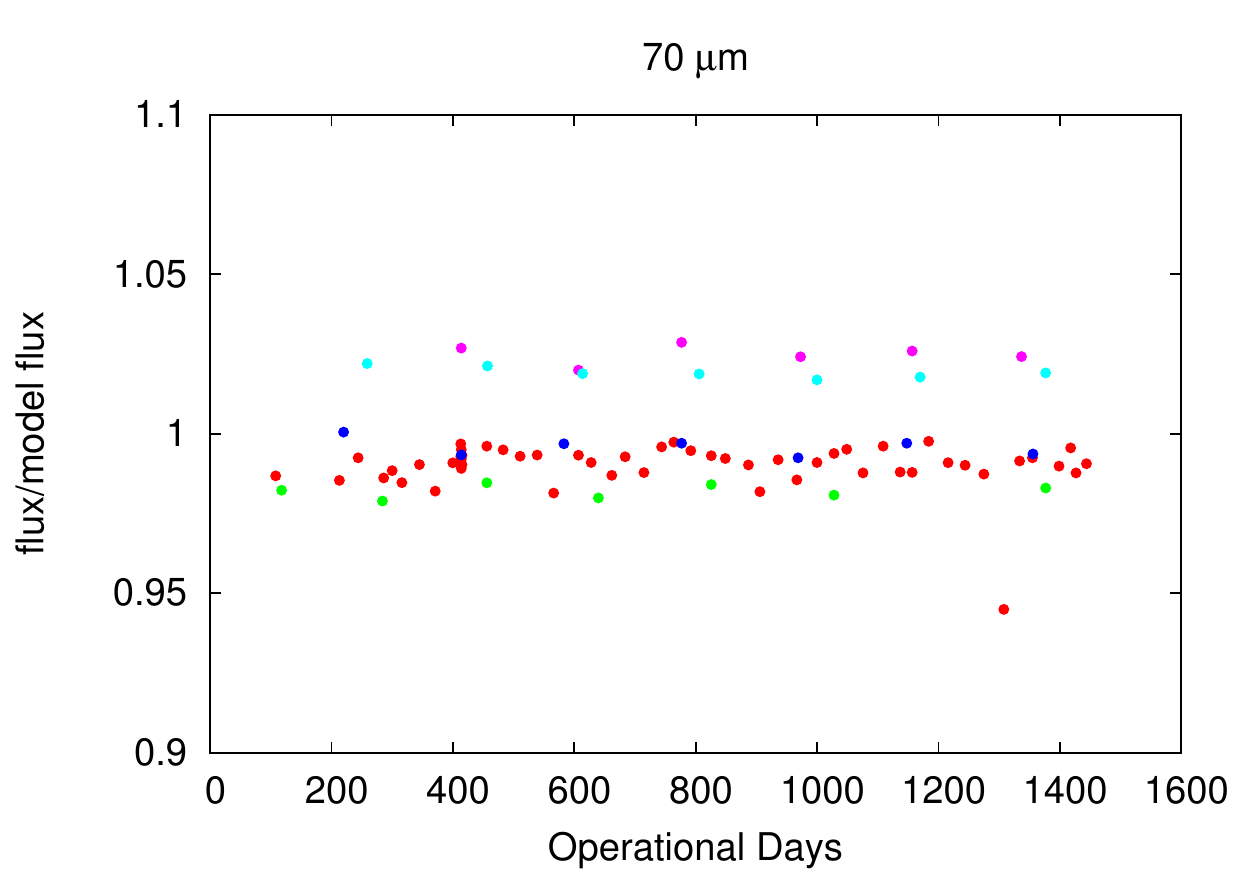}
\includegraphics[width=0.49\textwidth]{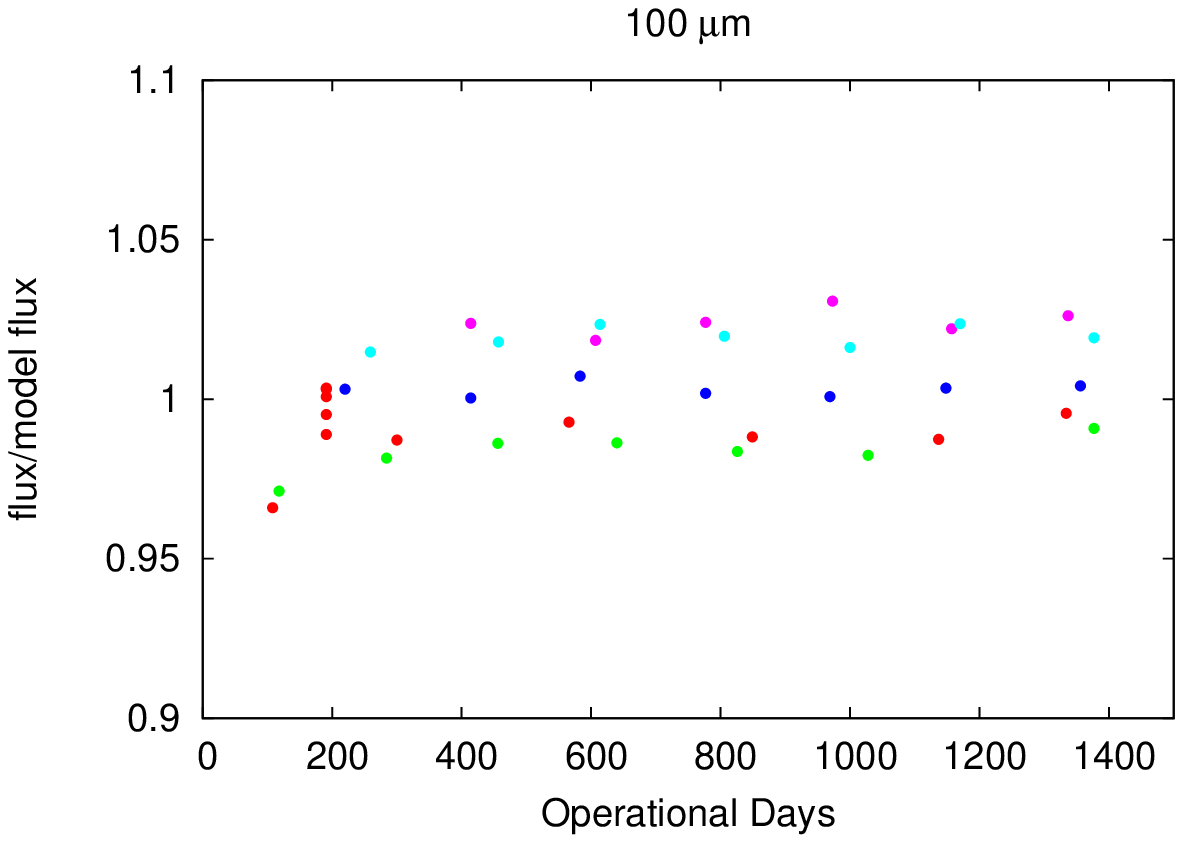}
\includegraphics[width=0.49\textwidth]{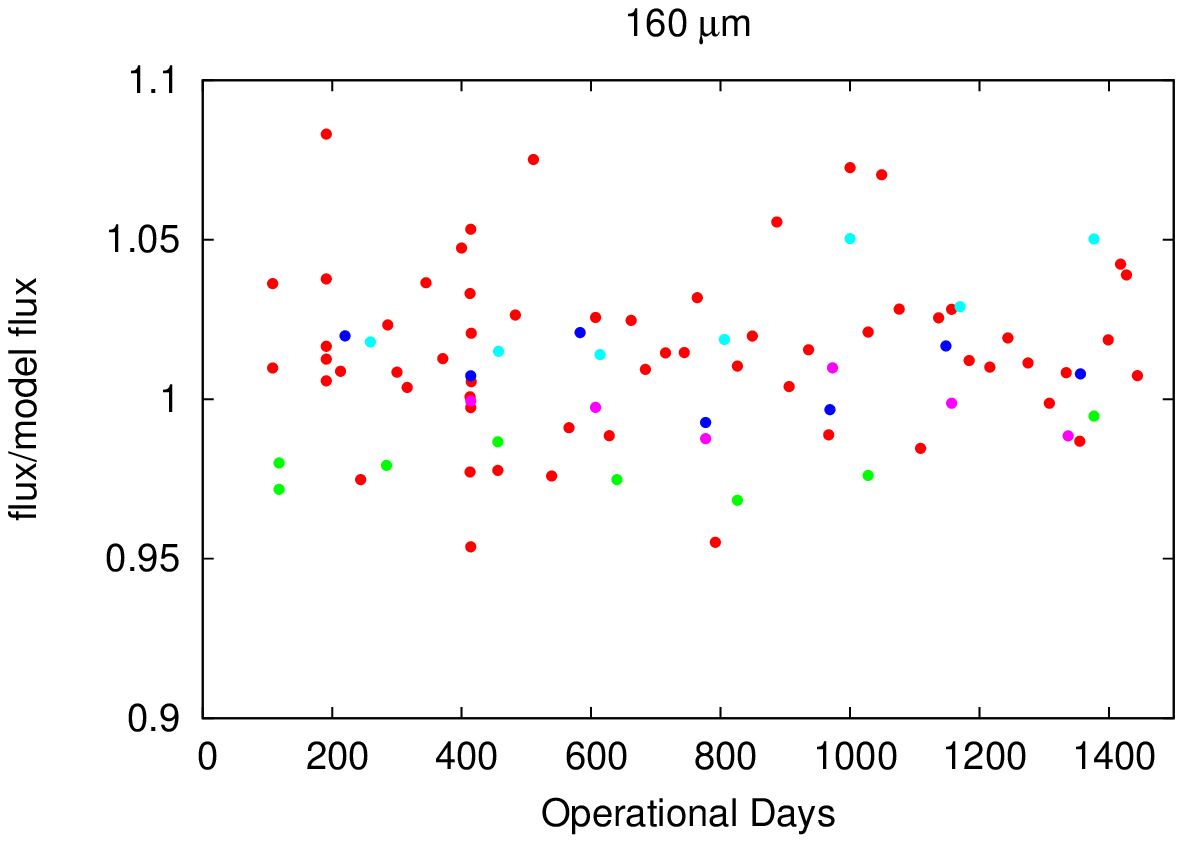}
\caption{Measured over model flux ratios of the five fiducial standards at 70 $\mu$m (upper left), 100 $\mu$m (upper right) and 160 $\mu$m (lower left) after applying the corrections for the evaporator temperature and the main mirror flux.}
\label{fig:oldstd_EVC_tflux}
\end{figure}

Figure \ref{fig:oldstd_EVC_tflux} shows again the same as Figure \ref {fig:oldstd} and \ref{fig:oldstd_EVC} after applying the correction for the evaporator temperature and the telescope flux. The trend with time observable especially in the blue band (left panel) in Figure \ref{fig:oldstd_EVC} disappears.

\begin{figure}

\includegraphics[width=0.75\textwidth]{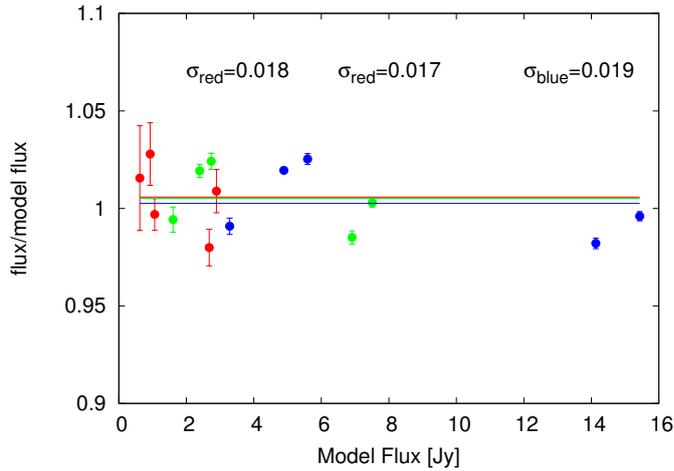}
\caption{Final result after applying the corrections. Red dots: 160 $\mu$m; green dots: 100 $\mu$m; blue dots: 70 $\mu$m. The red, green and blue lines represent the average values at  160 $\mu$m, 100 $\mu$m, and 70 $\mu$m respectively.}
\label{fig:finalres_corr}
\end{figure}

Finally, in Figure \ref{fig:finalres_corr} we reproduce Figure \ref{fig:finalres} using the corrected photometry. Note the decrease of the size of  the error bars in the blue and green band in Figure \ref{fig:finalres_corr} with respect to Figure \ref{fig:finalres}. 

\section{Conclusions}
\label{sec8}
We presented a quality check of the flux calibration scheme of the PACS photometer implemented in HIPE. Using the latest available responsivity calibration file (FM 7) we found that the measured fluxes of the five fiducial stars are in very good agreement with the model fluxes confirming our calibration and data reduction procedures. However we realize that there is a clear discrepancy (3-4\%) between the calibrators with different spectral types. 
The intrinsic uncertainty of the photometry of the individual sources is around or slightly below 2\%.  Adding the total of 5\% uncertainty of the stellar models gives 7\% as the most conservative estimate for the final error of our absolute flux calibration.  Therefore the overall uncertainty of the calibration is dominated by the models. We do not see any systematic wavelength dependence in the calibration.
The intrinsic accuracy of the photometry is affected by two instrumental effects. We have shown that these effects can be calibrated out in the blue and green bands bringing the intrinsic accuracy in these bands down to around 0.5\%.  It is important to note that these result are closely related to the way the observations were reduced. Different settings in the reduction scripts (for the masks, the deglitching, the high-pass filter widths, etc.) might produce slightly different fluxes. If one wants to be sure that the calibration is 100\% correct, then one has to take all OBSIDs of the 5 fiducial stars as listed in Tables  \ref{tbl:aBoo}, \ref{tbl:aCet}, \ref{tbl:aTau}, \ref{tbl:bAnd}, and \ref{tbl:gDra} and re-process the data with his/her own reduction script. Based on the new flux/model ratios one would have to adjust the fluxes of the science observations of interest. These adjustments are expected to be on a few percent level.


%
%

\begin{acknowledgements}
We would like to thank the anonymous referee for the comments and suggestions that significantly improved the manuscript. Z. Balog, H. Linz and M. Nielbock are funded by the Deutsches Zentrum f\"ur Luft- und Raumfahrt e. V.
\end{acknowledgements}

\bibliographystyle{spmpsci}      


\clearpage
\appendix
\section*{Appendix}
\label{sec:appendix}
\begin{table}[h!tb]
\begin{center}
\caption{PACS photometer observation details for $\alpha$\,Boo (HR\,5340; HD\,124897; HIP\,87833; Arcturus).}
  \label{tbl:aBoo}
\begin{tabular}{clccrrrl}
\noalign{\smallskip}
\hline
\noalign{\smallskip}
   &        & filter & scan-angles & \multicolumn{3}{c}{scan-legs} & notes/ \\ 
OD & OBSIDs & bands  & [deg]       & len [$^{\prime}$] & sep [$^{\prime \prime}$] & no.\ & remarks \\
\noalign{\smallskip}
\hline
\noalign{\smallskip}
220 & 1342188245, 1342188246 & b/r & 63/117  & 4.0 & 4.0 &  8 & \\
    & 1342188247, 1342188248 & g/r & 63/117  & 4.0 & 4.0 &  8 & \\
414 & 1342199603, 1342199604 & b/r & 70/110  & 3.0 & 4.0 & 10 & \\
    & 1342199606, 1342199607 & g/r & 70/110  & 3.0 & 4.0 & 10 & \\
583 & 1342211280, 1342211281 & b/r & 70/110  & 3.0 & 4.0 & 10 & \\
    & 1342211283, 1342211284 & g/r & 70/110  & 3.0 & 4.0 & 10 & \\
777 & 1342223345, 1342223346 & b/r & 70/110 & 3.0 & 4.0 & 10 & \\
    & 1342223348, 1342223349 & g/r & 70/110 & 3.0 & 4.0 & 10 & \\
969 & 1342236968, 1342236969 & b/r & 70/110 & 3.0 & 4.0 & 10 & \\
       & 1342236965, 1342236966 & g/r & 70/110 & 3.0 & 4.0 & 10 & \\
1148 & 1342247705, 1342247706 & b/r & 70/110 & 3.0 & 4.0 & 10 & \\
         & 1342247702, 1342247703 & g/r & 70/110 & 3.0 & 4.0 & 10 & \\
1356 & 1342262519, 1342262520 & b/r & 70/110& 3.0 & 4.0 & 10 & \\
          & 1342262516, 1342262517 & g/r & 70/110 & 3.0 & 4.0 & 10 & \\
\noalign{\smallskip}
\hline
\noalign{\smallskip}
\end{tabular}
\end{center}
\end{table}

\begin{table}[h!tb]
\begin{center}
\caption{PACS photometer observation details for $\alpha$\,Cet (HR\,911; HD\,18884; HIP\,14135; Menkar).}
  \label{tbl:aCet}
\begin{tabular}{clccrrrl}
\noalign{\smallskip}
\hline
\noalign{\smallskip}
   &        & filter & scan-angles & \multicolumn{3}{c}{scan-legs} & notes/ \\ 
OD & OBSIDs & bands  & [deg]       & len [$^{\prime}$] & sep [$^{\prime \prime}$] & no & remarks \\
\noalign{\smallskip}
\hline
\noalign{\smallskip}
259 & 1342189824, 1342189825 & b/r  &  63/117   &  4.0 &  4.0 &  8  & \\
    & 1342189827, 1342189828 & g/r  &  63/117	&  4.0 &  4.0 &  8  & \\
457 & 1342203030, 1342203031 & b/r  &  70/110	&  3.0 &  4.0 & 10  & \\
    & 1342203033, 1342203034 & g/r  &  70/110	&  3.0 &  4.0 & 10  & \\
614 & 1342212856, 1342212857 & b/r  &  70/110	&  3.0 &  4.0 & 10  & \\
    & 1342212853, 1342212854 & g/r  &  70/110	&  3.0 &  4.0 & 10  & \\
806 & 1342224927, 1342224928 & b/r & 70/110 & 3.0 & 4.0 & 10 & \\
    & 1342224930, 1342224931 & g/r & 70/110 & 3.0 & 4.0 & 10 & \\
1000 & 1342238782, 1342238783 & b/r & 70/110 & 3.0 & 4.0 & 10 & \\
    & 1342238779, 1342238780 & g/r & 70/110 & 3.0 & 4.0 & 10 & \\
1170 & 1342248722, 1342248723 & b/r & 70/110 & 3.0 & 4.0 & 10 & \\
    & 1342248719, 1342248720 & g/r & 70/110 & 3.0 & 4.0 & 10 & \\
1377 & 1342263910, 1342263911 & b/r & 70/110 & 3.0 & 4.0 & 10 & \\
    & 1342263907, 1342263908 & g/r & 70/110 & 3.0 & 4.0 & 10 & \\
\noalign{\smallskip}
\hline
\noalign{\smallskip}
\end{tabular}
\end{center}
\end{table}


\begin{table}[h!tb]
\begin{center}
\caption{PACS photometer observation details for $\alpha$\,Tau (HR\,1457; HD\,29139; HIP\,21421; Aldebaran).}
  \label{tbl:aTau}
\begin{tabular}{clccrrrl}
\noalign{\smallskip}
\hline
\noalign{\smallskip}
   &        & filter & scan-angles & \multicolumn{3}{c}{scan-legs} & notes/ \\ 
OD & OBSIDs & bands  & [deg]       & len [$^{\prime}$] & sep [$^{\prime \prime}$] & no & remarks \\
\noalign{\smallskip}
\hline
\noalign{\smallskip}
118 & 1342183532, 1342183533 & b/r &  45/135 &   5.0 & 51.0 &  4 & very low coverage \\
    & 1342183534, 1342183535 & g/r &  45/135 &   5.0 & 51.0 &  4 & very low coverage \\
118 & 1342183538             & b/r &  63     &  10.0 &  3.0 & 15 & no cross-scan \\
    & 1342183541             & g/r &  63     &  10.0 &  3.0 & 15 & no cross-scan \\
284 & 1342190947, 1342190948 & b/r &  70/110 &   2.5 &  4.0 & 10 & \\ 
    & 1342190944, 1342190945 & g/r &  70/110 &   2.5 &  4.0 & 10 & \\ 
456 & 1342202961, 1342202962 & b/r &  70/110 &   2.5 &  4.0 & 10 & \\ 
    & 1342202958, 1342202959 & g/r &  70/110 &   2.5 &  4.0 & 10 & \\ 
640 & 1342214211, 1342214212 & b/r & 70/110 & 3.0 & 4.0 & 10 & \\
 & 1342214214, 1342214215 & g/r & 70/110 & 3.0 & 4.0 & 10 & \\
826 & 1342226740, 1342226741 & b/r & 70/110 & 3.0 & 4.0 & 10 & \\
 & 1342226743, 1342226744 & g/r & 70/110 & 3.0 & 4.0 & 10 & \\
1028 & 1342240755, 1342240756 & b/r & 70/110 & 3.0 & 4.0 & 10 & \\
 & 1342240758, 1342240759 & g/r & 70/110 & 3.0 & 4.0 & 10 & \\
1377 & 1342263918, 1342263919 & b/r & 70/110 & 3.0 & 4.0 & 10 & \\
 & 1342263915, 1342263916 & g/r & 70/110 & 3.0 & 4.0 & 10 & \\

\noalign{\smallskip}
\hline
\noalign{\smallskip}
\end{tabular}
\end{center}
\end{table}


\begin{table}[h!tb]
\begin{center}
\caption{PACS photometer observation details for $\beta$\,And (HR\,337; HD\,6860; HIP\,5447; Mirach).}
  \label{tbl:bAnd}
\begin{tabular}{clccrrrl}
\noalign{\smallskip}
\hline
\noalign{\smallskip}
   &        & filter & scan-angles & \multicolumn{3}{c}{scan-legs} & notes/ \\ 
OD & OBSIDs & bands  & [deg]       & len [$^{\prime}$] & sep [$^{\prime \prime}$] & no & remarks \\
\noalign{\smallskip}
\hline
\noalign{\smallskip}
414 & 1342199609, 1342199610 & b/r &  70/110 &	3.0 &  4.0 & 10 & \\
    & 1342199612, 1342199613 & g/r &  70/110 &	3.0 &  4.0 & 10 & \\
607 & 1342212507, 1342212508 & b/r &  70/110 &	3.0 &  4.0 & 10 & \\
    & 1342212504, 1342212505 & g/r &  70/110 &	3.0 &  4.0 & 10 & \\
777 & 1342223335, 1342223336 & b/r & 70/110 & 3.0 & 4.0 & 10 & \\
 & 1342223338, 1342223339 & g/r & 70/110 & 3.0 & 4.0 & 10 & \\
973 & 1342237164, 1342237165 & b/r & 70/110 & 3.0 & 4.0 & 10 & \\
 & 1342237161, 1342237162 & g/r & 70/110 & 3.0 & 4.0 & 10 & \\
1157 & 1342248035, 1342248036 & b/r & 70/110 & 3.0 & 4.0 & 10 & \\
 & 1342248032, 1342248033 & g/r & 70/110 & 3.0 & 4.0 & 10 & \\
1337 & 1342259260, 1342259261 & b/r & 70/110 & 3.0 & 4.0 & 10 & \\
 & 1342259257, 1342259258 & g/r & 70/110 & 3.0 & 4.0 & 10 & \\
\noalign{\smallskip}
\hline
\noalign{\smallskip}
\end{tabular}
\end{center}
\end{table}


\begin{table}[h!tb]
\begin{center}
\caption{PACS photometer observation details for $\gamma$\,Dra (HR\,6705; HD\,164058; HIP\,87833; Etamin).}
  \label{tbl:gDra}
\begin{tabular}{clccrrrl}
\noalign{\smallskip}
\hline
\noalign{\smallskip}
   &        & filter & scan-angles & \multicolumn{3}{c}{scan-legs} & notes/ \\ 
OD & OBSIDs & bands  & [deg]       & len [$^{\prime}$] & sep [$^{\prime \prime}$] & no & remarks \\
\noalign{\smallskip}
\hline
\noalign{\smallskip}
108 & 1342182985, 1342182987 &  b/r  & 45/135  &  5.0 & 51.0 &  4 & very low coverage\\
108 & 1342182997, 1342182980 &  g/r  & 45/135  &  5.0 & 51.0 &  4 & very low coverage\\
108 &             1342182986 &  b/r &     45   & 30.0 &  4.0 & 15 & no cross-scan \\
    &             1342182981 &  g/r &     45   & 30.0 &  4.0 & 15 & no cross-scan \\
191 & 1342187147, 1342187148 &  g/r &	63/117  &  3.9 & 5.0 &  8 &  \\
    & 1342187149, 1342187150 &  g/r &	63/117  &  3.9 & 4.0 &  8 &  \\
    & 1342187151, 1342187152 &  g/r &	63/117  &  3.0 & 4.0 &  8 &  \\
    & 1342187153, 1342187154 &  g/r &	63/117  &  3.9 & 2.0 & 16 &  \\
    & 1342187155, 1342187156 &  g/r &	63/117  &  3.0 & 2.0 & 16 &  \\
213 & 1342188070, 1342188071 &  b/r &	 63/117  &  4.0 & 4.0 &  8 & \\ 
244 & 1342189187, 1342189187 &  b/r &	 63/117  &  4.0 & 4.0 &  8 & \\ 
286 & 1342191125, 1342191126 &  b/r &	 70/110  &  2.5 & 4.0 & 10 & \\ 
300 & 1342191958, 1342191959 &  b/r &	 70/110  &  2.5 & 4.0 & 10 & \\ 
    & 1342191961, 1342191962 &  g/r &	 70/110  &  2.5 & 4.0 & 10 & \\ 
316 & 1342192780, 1342192781 &  b/r &	 70/110  &  2.5 & 4.0 & 10 & \\  
345 & 1342195483, 1342195484 &  b/r &	 70/110  &  2.5 & 4.0 & 10 & \\  
371 & 1342196730, 1342196731 &  b/r &	 70/110  &  2.5 & 4.0 & 10 & \\  
400 & 1342198499, 1342198500 &  b/r &	 70/110  &  3.0 & 4.0 & 10 & \\  
413 & 1342199481, 1342199482 &  b/r &	 70/110  &  3.0 & 4.0 & 10 & \\  
    & 1342199512, 1342199513 &  b/r &	 70/110  &  3.0 & 4.0 & 10 & \\  
    & 1342199526, 1342199527 &  b/r &	 70/110  &  3.0 & 4.0 & 10 & \\  
414 & 1342199600, 1342199601 &  b/r &	 70/110  &  3.0 & 4.0 & 10 & \\  
    & 1342199639, 1342199640 &  b/r &	 70/110  &  3.0 & 4.0 & 10 & \\  
    & 1342199655, 1342199656 &  b/r &	 70/110  &  3.0 & 4.0 & 10 & \\  
415 & 1342199707, 1342199708 &  b/r &	 70/110  &  3.0 & 4.0 & 10 & \\  
    & 1342199717, 1342199718 &  b/r &	 70/110  &  3.0 & 4.0 & 10 & \\  
456 & 1342202942, 1342202943 &  b/r &	 70/110  &  3.0 & 4.0 & 10 & \\  
483 & 1342204209, 1342204210 &  b/r &	 70/110  &  3.0 & 4.0 & 10 & \\  
511 & 1342206001, 1342206002 &  b/r &	 70/110  &  3.0 & 4.0 & 10 & \\  
539 & 1342208971, 1342208972 &  b/r &	 70/110  &  3.0 & 4.0 & 10 & \\  
566 & 1342210582, 1342210583 &  b/r &	 70/110  &  3.0 & 4.0 & 10 & \\  
    & 1342210584, 1342210585 &  g/r &	 70/110  &  3.0 & 4.0 & 10 & \\  
607 & 1342212494, 1342212495 &  b/r &	 70/110  &  3.0 & 4.0 & 10 & \\  
628 & 1342213588,1342213589 & b/r & 70/110 & 3.0 & 4.0 & 10 & \\
662 & 1342215374,1342215375 & b/r & 70/110 & 3.0 & 4.0 & 10 & \\
684 & 1342217404,1342217405 & b/r & 70/110 & 3.0 & 4.0 & 10 & \\
715 & 1342220823,1342220824 & b/r & 70/110 & 3.0 & 4.0 & 10 & \\
744 & 1342221811,1342221812 & b/r & 70/110 & 3.0 & 4.0 & 10 & \\
764 & 1342222756, 1342222757 & b/r & 70/110 & 3.0 & 4.0 & 10 & \\
792 & 1342224229, 1342224230 & b/r & 70/110 & 3.0 & 4.0 & 10 & \\
826 & 1342226712, 1342226713 & b/r & 70/110 & 3.0 & 4.0 & 10 & \\
849 & 1342228388, 1342228389 & b/r & 70/110 & 3.0 & 4.0 & 10 & \\
 & 1342228391, 1342228392 & g/r & 70/110 & 3.0 & 4.0 & 10 & \\
887 & 1342231097, 1342231098 & b/r & 70/110 & 3.0 & 4.0 & 10 & \\
906 & 1342231899, 1342231900 & b/r & 70/110 & 3.0 & 4.0 & 10 & \\
936 & 1342234214, 1342234215 & b/r & 70/110 & 3.0 & 4.0 & 10 & \\
967 & 1342237975, 1342237976 & b/r & 70/110 & 3.0 & 4.0 & 10 & \\
1000 & 1342238772, 1342238773 & b/r & 70/110 & 3.0 & 4.0 & 10 & \\
1028 & 1342240699, 1342240700 & b/r & 70/110 & 3.0 & 4.0 & 10 & \\
1049 & 1342242557, 1342242558 & b/r & 70/110 & 3.0 & 4.0 & 10 & \\
1076 & 1342244900, 1342244901 & b/r & 70/110 & 3.0 & 4.0 & 10 & \\
1109 & 1342246181, 1342246182 & b/r & 70/110 & 3.0 & 4.0 & 10 & \\
1137 & 1342247335, 1342247336 & b/r & 70/110 & 3.0 & 4.0 & 10 & \\
 & 1342247338, 1342247339 & g/r & 70/110 & 3.0 & 4.0 & 10 & \\
1157 & 1342248038, 1342248039 & b/r & 70/110 & 3.0 & 4.0 & 10 & \\
1184 & 1342249293, 1342249294 & b/r & 70/110 & 3.0 & 4.0 & 10 & \\
1216 & 1342250856, 1342250857 & b/r & 70/110 & 3.0 & 4.0 & 10 & \\
1244 & 1342252805, 1342252806 & b/r & 70/110 & 3.0 & 4.0 & 10 & \\
1275 & 1342254723, 1342254724 & b/r & 70/110 & 3.0 & 4.0 & 10 & \\
1308 & 1342256959, 1342256960 & b/r & 70/110 & 3.0 & 4.0 & 10 & \\
1334 & 1342258831, 1342258832 & b/r & 70/110 & 3.0 & 4.0 & 10 & \\
 & 1342258834, 1342258835 & g/r & 70/110 & 3.0 & 4.0 & 10 & \\
1355 & 1342262225, 1342262226 & b/r & 70/110 & 3.0 & 4.0 & 10 & \\
1399 & 1342267291, 1342267292 & b/r & 70/110 & 3.0 & 4.0 & 10 & \\
1418 & 1342268966, 1342268967 & b/r & 70/110 & 3.0 & 4.0 & 10 & \\
1427 & 1342269812, 1342269813 & b/r & 70/110 & 3.0 & 4.0 & 10 & \\
1444 & 1342271000, 1342271001 & b/r & 70/110 & 3.0 & 4.0 & 10 & \\
\noalign{\smallskip}
\hline
\noalign{\smallskip}
\end{tabular}
\end{center}
\end{table}

\begin{table}[h!tb]
\begin{center}
\caption{PACS photometery of $\alpha$ Tau.  "with EVC" - fluxes taking into account the correction for the evaporator temperature (see Section 6.1); "without EVC" - fluxes without taking into account the correction for the evaporator temperature. This is also valid for Table \ref{tbl:alphaBoo_phot},\ref{tbl:alphaCet_phot},\ref{tbl:betaAnd_phot},\ref{tbl:gammaDra_phot}}
  \label{tbl:alphaTau_phot}
\begin{tabular}{|c|cccc|cccc|cccc|}
\noalign{\smallskip}
\hline
   OD&      \multicolumn{12}{c|}{filter} \\ 
\hline
 & \multicolumn{4}{|c|}{blue (70 $\mu$m)} &  \multicolumn{4}{c|}{green (100 $\mu$m)}&  \multicolumn{4}{c|}{red (160 $\mu$m)}\\
 \hline
 & \multicolumn{2}{c}{with EVC} &  \multicolumn{2}{c|}{without EVC}& \multicolumn{2}{c}{with EVC} &  \multicolumn{2}{c|}{without EVC}&\multicolumn{2}{c}{with EVC} &  \multicolumn{2}{c|}{without EVC}\\
 \hline
 & flux[Jy]& error[Jy]& flux[Jy]& error[Jy]& flux[Jy]& error[Jy]& flux[Jy]& error[Jy]& flux[Jy]& error[Jy]& flux[Jy]& error[Jy]\\
118 & 14.281 & 0.011 & 14.281 & 0.011 & 6.997 & 0.022 & 6.993 & 0.022 & 2.818 & 0.035 & 2.817 & 0.035\\
284 & 14.128 & 0.006 & 14.128 & 0.006 & 7.020 & 0.006 & 7.020 & 0.006 & 2.817 & 0.027 & 2.817 & 0.027\\
456 & 14.122 & 0.007 & 14.122 & 0.007 & 7.019 & 0.002 & 7.054 & 0.002 & 2.828 & 0.031 & 2.841 & 0.031\\
640 & 13.858 & 0.005 & 13.858 & 0.005 & 6.907 & 0.010 & 7.030 & 0.010 & 2.758 & 0.029 & 2.797 & 0.029\\
826 & 13.642 & 0.003 & 13.642 & 0.003 & 6.740 & 0.007 & 7.020 & 0.007 & 2.695 & 0.037 & 2.786 & 0.039\\
1028 & 13.801 & 0.009 & 13.801 & 0.009 & 6.864 & 0.002 & 6.983 & 0.002 & 2.766 & 0.028 & 2.797 & 0.028\\
1377 & 13.827 & 0.006 & 13.827 & 0.006 & 6.963 & 0.006 & 7.027 & 0.006 & 2.823 & 0.021 & 2.848 & 0.021\\
\hline
\noalign{\smallskip}
\end{tabular}
\end{center}
\end{table}

\begin{table}[h!tb]
\begin{center}
\caption{PACS photometery of $\alpha$ Boo}
  \label{tbl:alphaBoo_phot}
\begin{tabular}{|c|cccc|cccc|cccc|}
\noalign{\smallskip}
\hline
  OD &      \multicolumn{12}{c|}{filter} \\ 
\hline
 & \multicolumn{4}{|c|}{blue (70 $\mu$m)} &  \multicolumn{4}{c|}{green (100 $\mu$m)}&  \multicolumn{4}{c|}{red (160 $\mu$m)}\\
 \hline
 & \multicolumn{2}{c}{with EVC} &  \multicolumn{2}{c|}{without EVC}& \multicolumn{2}{c}{with EVC} &  \multicolumn{2}{c|}{without EVC}&\multicolumn{2}{c}{with EVC} &  \multicolumn{2}{c|}{without EVC}\\
 \hline
 & flux[Jy]& error[Jy]& flux[Jy]& error[Jy]& flux[Jy]& error[Jy]& flux[Jy]& error[Jy]& flux[Jy]& error[Jy]& flux[Jy]& error[Jy]\\
 220 & 15.839 & 0.007 & 15.817 & 0.007 & 7.828 & 0.009 & 7.818 & 0.009 & 3.181 & 0.044 & 3.177 & 0.044\\
414 & 15.593 & 0.013 & 15.672 & 0.013 & 7.752 & 0.010 & 7.791 & 0.010 & 3.125 & 0.031 & 3.138 & 0.031\\
583 & 15.655 & 0.004 & 15.634 & 0.004 & 7.812 & 0.008 & 7.803 & 0.007 & 3.167 & 0.058 & 3.163 & 0.058\\
777 & 15.197 & 0.010 & 15.651 & 0.011 & 7.521 & 0.006 & 7.781 & 0.006 & 3.003 & 0.028 & 3.088 & 0.029\\
969 & 15.302 & 0.001 & 15.475 & 0.001 & 7.652 & 0.009 & 7.726 & 0.009 & 3.053 & 0.050 & 3.082 & 0.050\\
1148 & 15.438 & 0.007 & 15.574 & 0.007 & 7.710 & 0.010 & 7.774 & 0.010 & 3.135 & 0.033 & 3.159 & 0.034\\
1356 & 15.228 & 0.003 & 15.418 & 0.003 & 7.655 & 0.001 & 7.735 & 0.001 & 3.082 & 0.036 & 3.113 & 0.036\\
\hline
\noalign{\smallskip}
\end{tabular}
\end{center}
\end{table}

\begin{table}[h!tb]
\begin{center}
\caption{PACS photometery of $\alpha$ Cet}
  \label{tbl:alphaCet_phot}
\begin{tabular}{|c|cccc|cccc|cccc|}
\noalign{\smallskip}
\hline
   OD&      \multicolumn{12}{c|}{filter} \\ 
\hline
 & \multicolumn{4}{|c|}{blue (70 $\mu$m)} &  \multicolumn{4}{c|}{green (100 $\mu$m)}&  \multicolumn{4}{c|}{red (160 $\mu$m)}\\
 \hline
 & \multicolumn{2}{c}{with EVC} &  \multicolumn{2}{c|}{without EVC}& \multicolumn{2}{c}{with EVC} &  \multicolumn{2}{c|}{without EVC}&\multicolumn{2}{c}{with EVC} &  \multicolumn{2}{c|}{without EVC}\\
 \hline
 & flux[Jy]& error[Jy]& flux[Jy]& error[Jy]& flux[Jy]& error[Jy]& flux[Jy]& error[Jy]& flux[Jy]& error[Jy]& flux[Jy]& error[Jy]\\
 259 & 5.071 & 0.002 & 5.104 & 0.002 & 2.498 & 0.006 & 2.514 & 0.006 & 1.009 & 0.009 & 1.015 & 0.009\\
457 & 5.055 & 0.005 & 5.094 & 0.005 & 2.503 & 0.003 & 2.522 & 0.003 & 1.006 & 0.004 & 1.013 & 0.004\\
614 & 5.036 & 0.006 & 5.058 & 0.006 & 2.515 & 0.003 & 2.525 & 0.003 & 1.004 & 0.040 & 1.008 & 0.040\\
806 & 5.036 & 0.005 & 5.060 & 0.005 & 2.509 & 0.006 & 2.521 & 0.006 & 1.012 & 0.016 & 1.016 & 0.016\\
1000 & 5.011 & 0.004 & 5.024 & 0.004 & 2.496 & 0.004 & 2.502 & 0.004 & 1.041 & 0.027 & 1.043 & 0.027\\
1170 & 5.027 & 0.007 & 5.031 & 0.007 & 2.523 & 0.003 & 2.525 & 0.003 & 1.024 & 0.014 & 1.025 & 0.014\\
1377 & 4.967 & 0.005 & 5.012 & 0.005 & 2.482 & 0.004 & 2.504 & 0.004 & 1.034 & 0.024 & 1.042 & 0.025\\
\hline
\noalign{\smallskip}
\end{tabular}
\end{center}
\end{table}

\begin{table}[h!tb]
\begin{center}
\caption{PACS photometery of $\beta$ And}
  \label{tbl:betaAnd_phot}
\begin{tabular}{|c|cccc|cccc|cccc|}
\noalign{\smallskip}
\hline
   OD&      \multicolumn{12}{c|}{filter} \\ 
\hline
 & \multicolumn{4}{|c|}{blue (70 $\mu$m)} &  \multicolumn{4}{c|}{green (100 $\mu$m)}&  \multicolumn{4}{c|}{red (160 $\mu$m)}\\
 \hline
 & \multicolumn{2}{c}{with EVC} &  \multicolumn{2}{c|}{without EVC}& \multicolumn{2}{c}{with EVC} &  \multicolumn{2}{c|}{without EVC}&\multicolumn{2}{c}{with EVC} &  \multicolumn{2}{c|}{without EVC}\\
 \hline
 & flux[Jy]& error[Jy]& flux[Jy]& error[Jy]& flux[Jy]& error[Jy]& flux[Jy]& error[Jy]& flux[Jy]& error[Jy]& flux[Jy]& error[Jy]\\
414 & 5.842 & 0.007 & 5.872 & 0.007 & 2.891 & 0.005 & 2.906 & 0.005 & 1.139 & 0.024 & 1.144 & 0.024\\
607 & 5.741 & 0.006 & 5.792 & 0.006 & 2.849 & 0.004 & 2.873 & 0.004 & 1.125 & 0.017 & 1.134 & 0.016\\
777 & 5.802 & 0.009 & 5.853 & 0.009 & 2.875 & 0.004 & 2.899 & 0.004 & 1.121 & 0.022 & 1.129 & 0.022\\
973 & 5.792 & 0.005 & 5.788 & 0.005 & 2.903 & 0.006 & 2.901 & 0.006 & 1.148 & 0.026 & 1.147 & 0.025\\
1157 & 5.776 & 0.005 & 5.805 & 0.005 & 2.870 & 0.007 & 2.884 & 0.008 & 1.134 & 0.016 & 1.139 & 0.016\\
1337 & 5.734 & 0.007 & 5.763 & 0.008 & 2.868 & 0.004 & 2.882 & 0.004 & 1.117 & 0.013 & 1.122 & 0.013\\
\hline
\noalign{\smallskip}
\end{tabular}
\end{center}
\end{table}

\begin{table}[h!tb]
\begin{center}
\caption{PACS photometery of $\gamma$ Dra}
  \label{tbl:gammaDra_phot}
\begin{tabular}{|c|cccc|cccc|cccc|}
\noalign{\smallskip}
\hline
   OD&      \multicolumn{12}{c|}{filter} \\ 
\hline
 & \multicolumn{4}{|c|}{blue (70 $\mu$m)} &  \multicolumn{4}{c|}{green (100 $\mu$m)}&  \multicolumn{4}{c|}{red (160 $\mu$m)}\\
 \hline
 & \multicolumn{2}{c}{with EVC} &  \multicolumn{2}{c|}{without EVC}& \multicolumn{2}{c}{with EVC} &  \multicolumn{2}{c|}{without EVC}&\multicolumn{2}{c}{with EVC} &  \multicolumn{2}{c|}{without EVC}\\
 \hline
 & flux[Jy]& error[Jy]& flux[Jy]& error[Jy]& flux[Jy]& error[Jy]& flux[Jy]& error[Jy]& flux[Jy]& error[Jy]& flux[Jy]& error[Jy]\\
108 & 3.313 & 0.012 & 3.331 & 0.012 & 1.606 & 0.008 & 1.615 & 0.008 & 0.694 & 0.018 & 0.697 & 0.018\\
191 & & & & & 1.659 & 0.008 & 1.667 & 0.008 & 0.722 & 0.018 & 0.725 & 0.019\\
191 & & & & & 1.662 & 0.006 & 1.671 & 0.006 & 0.678 & 0.011 & 0.681 & 0.011\\
191 & & & & & 1.639 & 0.007 & 1.647 & 0.007 & 0.692 & 0.014 & 0.695 & 0.015\\
191 & & & & & 1.663 & 0.004 & 1.672 & 0.004 & 0.675 & 0.016 & 0.678 & 0.017\\
191 & & & & & 1.649 & 0.005 & 1.658 & 0.005 & 0.670 & 0.006 & 0.673 & 0.006\\
213 & 3.302 & 0.006 & 3.315 & 0.006 & & & & & 0.675 & 0.019 & 0.675 & 0.019\\
244 & 3.318 & 0.011 & 3.333 & 0.011 & & & & & 0.649 & 0.008 & 0.651 & 0.008\\
286 & 3.281 & 0.004 & 3.307 & 0.004 & & & & & 0.679 & 0.018 & 0.683 & 0.018\\
300 & 3.294 & 0.004 & 3.313 & 0.004 & 1.629 & 0.003 & 1.639 & 0.003 & 0.669 & 0.010 & 0.673 & 0.010\\
316 & 3.274 & 0.002 & 3.304 & 0.002 & & & & & 0.666 & 0.017 & 0.671 & 0.017\\
345 & 3.298 & 0.002 & 3.323 & 0.002 & & & & & 0.688 & 0.033 & 0.693 & 0.033\\
371 & 3.303 & 0.003 & 3.294 & 0.003 & & & & & 0.679 & 0.015 & 0.677 & 0.015\\
400 & 3.307 & 0.008 & 3.323 & 0.008 & & & & & 0.697 & 0.020 & 0.700 & 0.021\\
413 & 3.336 & 0.007 & 3.331 & 0.007 & & & & & 0.670 & 0.024 & 0.670 & 0.024\\
413 & 3.336 & 0.006 & 3.345 & 0.006 & & & & & 0.690 & 0.015 & 0.691 & 0.015\\
413 & 3.308 & 0.002 & 3.323 & 0.002 & & & & & 0.651 & 0.016 & 0.654 & 0.016\\
414 & 3.323 & 0.005 & 3.339 & 0.005 & & & & & 0.636 & 0.008 & 0.638 & 0.008\\
414 & 3.301 & 0.003 & 3.320 & 0.003 & & & & & 0.701 & 0.028 & 0.705 & 0.028\\
414 & 3.307 & 0.006 & 3.329 & 0.006 & & & & & 0.664 & 0.008 & 0.668 & 0.008\\
415 & 3.310 & 0.005 & 3.333 & 0.005 & & & & & 0.679 & 0.010 & 0.683 & 0.010\\
415 & 3.295 & 0.003 & 3.324 & 0.003 & & & & & 0.668 & 0.016 & 0.673 & 0.016\\
456 & 3.321 & 0.005 & 3.337 & 0.006 & & & & & 0.650 & 0.025 & 0.653 & 0.025\\
483 & 3.327 & 0.003 & 3.329 & 0.003 & & & & & 0.685 & 0.016 & 0.685 & 0.016\\
511 & 3.325 & 0.005 & 3.323 & 0.005 & & & & & 0.718 & 0.019 & 0.718 & 0.019\\
539 & 3.329 & 0.009 & 3.322 & 0.009 & & & & & 0.653 & 0.012 & 0.651 & 0.012\\
566 & 3.248 & 0.005 & 3.277 & 0.005 & 1.630 & 0.004 & 1.645 & 0.004 & 0.655 & 0.014 & 0.660 & 0.014\\
607 & 3.283 & 0.004 & 3.310 & 0.004 & & & & & 0.677 & 0.014 & 0.682 & 0.014\\
628 & 3.283 & 0.004 & 3.303 & 0.004 & & & & & 0.654 & 0.012 & 0.658 & 0.012\\
662 & 3.273 & 0.007 & 3.290 & 0.007 & & & & & 0.679 & 0.018 & 0.682 & 0.019\\
684 & 3.295 & 0.003 & 3.311 & 0.003 & & & & & 0.670 & 0.012 & 0.673 & 0.012\\
715 & 3.279 & 0.002 & 3.298 & 0.002 & & & & & 0.674 & 0.021 & 0.677 & 0.021\\
744 & 3.302 & 0.007 & 3.322 & 0.007 & & & & & 0.673 & 0.027 & 0.677 & 0.027\\
764 & 3.220 & 0.006 & 3.329 & 0.007 & & & & & 0.669 & 0.007 & 0.689 & 0.007\\
792 & 3.200 & 0.002 & 3.320 & 0.002 & & & & & 0.618 & 0.020 & 0.638 & 0.020\\
826 & 3.285 & 0.004 & 3.311 & 0.004 & & & & & 0.670 & 0.020 & 0.674 & 0.021\\
849 & 3.275 & 0.003 & 3.302 & 0.003 & 1.621 & 0.008 & 1.634 & 0.008& 0.674 & 0.015 & 0.679 & 0.015\\
887 & 3.298 & 0.006 & 3.292 & 0.006 & & & & & 0.704 & 0.011 & 0.703 & 0.011\\
906 & 3.236 & 0.005 & 3.264 & 0.005 & & & & & 0.663 & 0.034 & 0.668 & 0.035\\
936 & 3.269 & 0.006 & 3.292 & 0.006 & & & & & 0.672 & 0.015 & 0.675 & 0.016\\
967 & 3.261 & 0.006 & 3.269 & 0.006 & & & & & 0.655 & 0.014 & 0.657 & 0.014\\
1000 & 3.283 & 0.005 & 3.288 & 0.005 & & & & & 0.712 & 0.019 & 0.713 & 0.019\\
1028 & 3.268 & 0.008 & 3.297 & 0.008 & & & & & 0.674 & 0.013 & 0.679 & 0.013\\
1049 & 3.302 & 0.006 & 3.305 & 0.006 & & & & & 0.712 & 0.011 & 0.713 & 0.011\\
1076 & 3.227 & 0.005 & 3.281 & 0.005 & & & & & 0.677 & 0.016 & 0.685 & 0.016\\
1109 & 3.283 & 0.006 & 3.305 & 0.006 & & & & & 0.651 & 0.017 & 0.655 & 0.017\\
1137 & 3.280 & 0.007 & 3.281 & 0.007 & 1.632 & 0.004 & 1.633 & 0.004 & 0.684 & 0.021 & 0.684 & 0.021\\
1157 & 3.264 & 0.005 & 3.280 & 0.005 & & & & & 0.683 & 0.026 & 0.686 & 0.026\\
1184 & 3.279 & 0.004 & 3.308 & 0.004 & & & & & 0.669 & 0.013 & 0.674 & 0.013\\
1216 & 3.254 & 0.003 & 3.283 & 0.003 & & & & & 0.667 & 0.029 & 0.672 & 0.029\\
1244 & 3.269 & 0.005 & 3.280 & 0.005 & & & & & 0.676 & 0.008 & 0.678 & 0.008\\
1275 & 3.237 & 0.004 & 3.265 & 0.004 & & & & & 0.667 & 0.013 & 0.672 & 0.013\\
1308 & 3.127 & 0.008 & 3.121 & 0.007 & & & & & 0.664 & 0.024 & 0.663 & 0.024\\
1334 & 3.244 & 0.004 & 3.275 & 0.004 & 1.622 & 0.004 & 1.639 & 0.004 & 0.664 & 0.010 & 0.669 & 0.010\\
1355 & 3.257 & 0.005 & 3.276 & 0.005 & & & & & 0.652 & 0.012 & 0.655 & 0.012\\
1399 & 3.253 & 0.004 & 3.269 & 0.004 & & & & & 0.674 & 0.042 & 0.676 & 0.042\\
1418 & 3.286 & 0.003 & 3.288 & 0.003 & & & & & 0.692 & 0.019 & 0.692 & 0.019\\
1427 & 3.239 & 0.008 & 3.262 & 0.008 & & & & & 0.686 & 0.021 & 0.690 & 0.021\\
1444 & 3.249 & 0.002 & 3.271 & 0.002 & & & & & 0.665 & 0.025 & 0.669 & 0.025\\
\hline
\noalign{\smallskip}
\end{tabular}
\end{center}
\end{table}

\end{document}